\documentclass[sigconf]{acmart}

\usepackage[utf8]{inputenc}
\usepackage[T1]{fontenc}
\usepackage{graphicx}
\usepackage{tabularx}
\usepackage{subcaption}
\usepackage[english]{babel}
\usepackage[figuresright]{rotating}
\usepackage{makecell}
\usepackage{array}
\usepackage{caption}
\usepackage{multirow}
\usepackage{url}
\usepackage{float}
\usepackage{hyperref}
\usepackage{rotating} 
\usepackage{array}
\usepackage{makecell}
\usepackage{tikz}
\usepackage{afterpage}
\usepackage{longtable,lscape}
\usepackage{enumitem}
\usepackage{color, colortbl}
\usepackage{makecell}
\usepackage{wrapfig}
\usepackage[linesnumbered]{algorithm2e}
\definecolor{Gray}{gray}{0.9}
\definecolor{LightGray}{gray}{0.6}
\hypersetup{colorlinks=true, citecolor={magenta}, filecolor=blue, linkcolor=black, urlcolor=blue}

\usepackage{xcolor}
\definecolor{green(munsell)}{rgb}{0.0, 0.66, 0.47}
\definecolor{cadmiumgreen}{rgb}{0.0, 0.42, 0.24}
\definecolor{cobalt}{rgb}{0.0, 0.28, 0.67}
\definecolor{amber(sae/ece)}{rgb}{1.0, 0.49, 0.0}
\newlength\MAX  

\newlength\BARSIZE  

\copyrightyear{2023}
\acmYear{2023}
\setcopyright{acmlicensed}\acmConference[CHI '23]{Proceedings of the 2023 CHI Conference on Human Factors in Computing Systems}{April 23--28, 2023}{Hamburg, Germany}
\acmBooktitle{Proceedings of the 2023 CHI Conference on Human Factors in Computing Systems (CHI '23), April 23--28, 2023, Hamburg, Germany}
\acmPrice{15.00}
\acmDOI{10.1145/3544548.3581190}
\acmISBN{978-1-4503-9421-5/23/04}

\makeatletter
\def\@fnsymbol#1{\ensuremath{\ifcase#1\or \dagger\or 
   \mathsection\or \mathparagraph\or \|\or **\or \dagger\dagger
   \or \ddagger\ddagger \else\@ctrerr\fi}}
\makeatother

\definecolor{Gray}{gray}{0.9}
\definecolor{Gray2}{gray}{0.8}

\begin{document}

\title{Complex Daily Activities, Country-Level Diversity, and Smartphone Sensing: A Study in Denmark, Italy, Mongolia, Paraguay, and UK}

\author{Karim Assi}
\authornote{Both authors contributed equally and are listed alphabetically.}
\orcid{0000-0002-8787-5058}
\affiliation{
\institution{EPFL} \country{Switzerland}}

\author{Lakmal Meegahapola}
\authornotemark[1]
\authornote{Corresponding authors are Lakmal Meegahapola (\url{lakmal.meegahapola@epfl.ch}) and Daniel Gatica-Perez (\url{daniel.gatica-perez@epfl.ch}).}
\orcid{0000-0002-5275-6585}
\affiliation{
\institution{Idiap Research Institute \& EPFL} \country{Switzerland}}

\author{William Droz}
\orcid{0000-0003-0379-2018}
\affiliation{\institution{Idiap Research Institute} \country{Switzerland}}

\author{Peter Kun}
\orcid{0000-0003-0778-7662}
\affiliation{\institution{Aalborg University} \country{Denmark}}

\author{Amalia de Götzen}
\orcid{0000-0001-7214-5856}
\affiliation{\institution{Aalborg University} \country{Denmark}}

\author{Miriam Bidoglia}
\orcid{0000-0002-1583-6551}
\affiliation{\institution{London School of Economics and Political Science, UK}\country{}}

\author{Sally Stares}
\orcid{0000-0003-4697-0347}
\affiliation{\institution{City, University of London}\country{UK}}

\author{George Gaskell}
\orcid{0000-0001-6135-9496}
\affiliation{\institution{London School of Economics and Political Science, UK}\country{}}

\author{Altangerel Chagnaa}
\orcid{0000-0003-2331-3045}
\affiliation{\institution{National University of Mongolia} \country{Mongolia}}

\author{Amarsanaa Ganbold}
\orcid{0000-0003-4335-6608}
\affiliation{\institution{National University of Mongolia} \country{Mongolia}}

\author{Tsolmon Zundui}
\orcid{0000-0002-2797-517X}
\affiliation{\institution{National University of Mongolia} \country{Mongolia}}

\author{Carlo Caprini}
\orcid{0000-0001-6609-1572}
\affiliation{\institution{U-Hopper} \country{Italy}}

\author{Daniele Miorandi}
\orcid{0000-0002-3089-977X}
\affiliation{\institution{U-Hopper} \country{Italy}}

\author{Jose Luis Zarza}
\orcid{0000-0003-0069-4287}
\affiliation{\institution{Universidad Católica "Nuestra Señora de la Asunción", Paraguay} \country{}}

\author{Alethia Hume}
\orcid{0000-0002-1874-1419}
\affiliation{\institution{Universidad Católica "Nuestra Señora de la Asunción", Paraguay} \country{}}

\author{Luca Cernuzzi}
\orcid{0000-0001-7803-1067}
\affiliation{\institution{Universidad Católica "Nuestra Señora de la Asunción", Paraguay}\country{}}

\author{Ivano Bison}
\orcid{0000-0002-9645-8627}
\affiliation{\institution{University of Trento} \country{Italy}}

\author{Marcelo Dario Rodas Britez}
\orcid{0000-0002-7607-7587}
\affiliation{\institution{University of Trento} \country{Italy}}

\author{Matteo Busso}
\orcid{0000-0002-3788-0203}
\affiliation{\institution{University of Trento} \country{Italy}}

\author{Ronald Chenu-Abente}
\orcid{0000-0002-1121-0287}
\affiliation{\institution{University of Trento} \country{Italy}}

\author{Fausto Giunchiglia}
\orcid{0000-0002-5903-6150}
\affiliation{\institution{University of Trento} \country{Italy}}

\author{Daniel Gatica-Perez}
\authornotemark[2]
\orcid{0000-0001-5488-2182}
\affiliation{\institution{Idiap Research Institute \& EPFL} \country{Switzerland}}

\renewcommand{\shortauthors}{Assi and Meegahapola et al.}
\renewcommand{\shorttitle}{Complex Daily Activities, Country-Level Diversity and Smartphone Sensing}

\begin{abstract}

%
%
Smartphones enable understanding human behavior with activity recognition to support people’s daily lives. Prior studies focused on using inertial sensors to detect simple activities (sitting, walking, running, etc.) and were mostly conducted in homogeneous populations within a country. However, people are more sedentary in the post-pandemic world with the prevalence of remote/hybrid work/study settings, making detecting simple activities less meaningful for context-aware applications. Hence, the understanding of \emph{(i)} how multimodal smartphone sensors and machine learning models could be used to detect complex daily activities that can better inform about people’s daily lives, and \emph{(ii)} how models generalize to unseen countries, is limited. We analyzed in-the-wild smartphone data and $\sim$216K self-reports from 637 college students in five countries (Italy, Mongolia, UK, Denmark, Paraguay). Then, we defined a 12-class complex daily activity recognition task and evaluated the performance with different approaches. We found that even though the generic multi-country approach provided an AUROC of 0.70, the country-specific approach performed better with AUROC scores in [0.79-0.89]. We believe that research along the lines of diversity awareness is fundamental for advancing human behavior understanding through smartphones and machine learning, for more real-world utility across countries.

\end{abstract}



\maketitle

\section{Introduction}

The field of activity recognition has gained substantial attention in recent years due to its usefulness in various domains, including healthcare \cite{straczkiewicz_systematic_2021}, sports \cite{zhuang_sport-related_2019}, transportation \cite{naseeb_activity_2020}, and human well-being \cite{meegahapola2020smartphone}. For instance, fitness-tracking mobile health applications enable users to access activity-specific metrics \cite{zhuang_sport-related_2019, siirtola_efficient_2011}. Similarly, smart home systems can make changes to the environment (e.g., lighting, temperature) based on the information gathered about people's activities \cite{makonin_smarter_2013, nef_evaluation_2015}. Context awareness, a key aspect of mobile phone user experience, is enabled with the integration of activity recognition \cite{wang_context_aware_music_rec, oh2015intelligent}.

Traditionally, sensor-based activity recognition relied on custom sensors attached to the body \cite{choudhury_mobile_2008}. While this approach is effective for small-scale studies, it is often challenging to scale up. The cost and maintenance required for these sensors can make them both expensive and obtrusive, reducing the motivation to use them. The alternative approach of using commercial wearables is not immune to these challenges, and these devices are often perceived as niche or abandoned after a short period of usage \cite{mercer_acceptance_2016, coorevits_rise_2016}. This is where the presence of smartphones comes in handy. In the United States, 85\% of adults and 96\% of young adults own a smartphone, making it easier to target a broader audience \cite{noauthor_mobile_2021}. Research in mobile sensing has revealed the potential of smartphone data for activity recognition \cite{straczkiewicz_systematic_2021, meegahapola2020smartphone}. The widespread ownership and unobtrusive nature of smartphones make them an attractive solution to traditional sensor-based activity recognition. However, there is still a need to understand how multiple sensing modalities in smartphones can be utilized for complex daily activity recognition. Additionally, the generalization of complex daily activity recognition models across different countries remains an under-explored area of research.

Recognizing complex daily activities is important. In the activity recognition literature, multiple types of activities have been considered, each at different granularity levels \cite{dernbach_simple_2012, saguna2013complex}. Coarse-grained or simple activities like walking, sitting, or cycling are repeated \emph{unitary} actions directly measurable from a proxy (e.g., inertial sensor unit). Fine-grained complex activities, or activities of daily living (ADL), are built on top of simple activities, but convey more specific contextual information \cite{wiener1990measuring, rai2012mining, saguna2013complex}. For example, eating, studying, working, and movie watching entail participants sitting. Such activities can not be measured by inertial sensor units alone \cite{boutonyour2022, meegahapola2020protecting, biel_bitesnbits_2018} and need a more holistic multimodal sensing approach that captures a wide range of contexts and behaviors that build on top of simple activities \cite{saguna2013complex}. Further, recognizing such complex daily activities could: \emph{(i)} allow tracking the digital well-being of individuals in a more fine-grained manner (e.g., providing a breakdown of time spent eating, resting, attending a lecture, and studying, instead of just sitting \cite{sewall2020psychosocial, boutonyour2022}); \emph{(ii)} provide context-aware user experiences and notifications by understanding user behavior better (e.g., not sending phone notifications when a person is studying or attending a lecture, suggesting products while a user is shopping \cite{mehrotra2017intelligent}); and \emph{(iii)} allow better content recommendation (e.g., recommending music based on the current daily activity such as working, studying, or shopping \cite{wang_context_aware_music_rec}), where complex activities can be more informative and valuable than simpler ones. However, even though inertial, location, or WiFi/Bluetooth data have been used separately for activity recognition \cite{rai2012mining, saguna2013complex}, prior work has not exhaustively studied complex daily activities by using multimodal smartphone sensing data.

The use of multimodal smartphone sensing data in machine learning models could provide a more comprehensive picture of complex daily activities when compared to using single modalities. This is especially relevant in light of the Covid-19 pandemic, which has brought about a significant shift in daily habits and activities \cite{stockwell_changes_2021, zheng_covid-19_2020}. The lockdown measures enforced to slow the spread of the virus resulted in a decrease in physical activity and an increase in sedentary behavior, particularly among young adults. This shift is evident in changes to smartphone use patterns \cite{ratan_smartphone_2021, saadeh_smartphone_2021, li_impact_2021}, which can impact the effectiveness of location-based activity recognition methods in a remote/hybrid work/study setting where individuals tend to remain sedentary for extended periods of time. Hence, the importance of inertial and location sensors as predictive features could diminish due to sedentary behavior. This underscores the importance of incorporating fine-grained multimodal sensing features to accurately characterize the complex daily activities of these emerging lifestyles through smartphones. However, there is currently little understanding of which smartphone sensing features are systematically useful in characterizing different complex daily activities.

Taking a few steps back, we can also consider the ``country'' dimension and its influence on smartphone usage. Country differences can affect smartphone usage in different world regions \cite{mathur_moving_2017}. For example, it could be socially frowned upon to take a call at a formal restaurant in Japan, while people in Europe could leave a movie theater to check their phone \cite{canton_cell_2012}. It has been shown that people in Japan tend to be more reticent than in Sweden about talking on the phone in public transportation or, more generally, about being loud in public \cite{phone_usage_jp_sw_usa}. Another study about smartphone addiction among young adults in 24 countries found that the rigidity of social norms and obligations highly influenced smartphone usage \cite{smartphone_addiction_24_countries}. In addition to how people use the phone, prior work also discussed how passively sensed behavioral data about people differ in many countries \cite{althoff2017large}. These differences across countries constitute a form of diversity, which is a growing area of interest in computing and AI research \cite{AICultures2023} \footnote{While we acknowledge that cultures can be multidimensional and exist in tension with each other and in plurality within the same country \cite{yuval2004gender}, some prior studies in mobile sensing, psychology, and sociology have used ``culture'' as a proxy to refer to the country of data collection \cite{phan_mobile_2022, khwaja_modeling_2019, van2004bias, he2012bias}. However, in this study, for consistency, we use ``country'' (a more specific geographic region) as the unit of analysis that could affect phone usage behavior and sensing data. We also used the term ``geographic'' rarely, when appropriate and when referring to regions (i.e., Europe).}. From a machine learning point-of-view, a diversified system contains more information and can better fit various environments \cite{gong_diversity_2019}. More generally, diversity-aware machine learning aims to improve the model's representational ability through various components such as input data, parameters, and outputs \cite{gong_diversity_2019}. Concretely, country-level, diversity-aware activity recognition should try to understand the effect of the country diversity of smartphone users, on inference model performance. However, the understanding of how country diversity affects the smartphone sensing pipeline (from collected data to model performance) is limited, as previous work aimed at quantifying such effects has been scarce \cite{khwaja_modeling_2019, meegahapola2020smartphone, phan_mobile_2022}, due to reasons including, but not limited to, logistical difficulties in conducting longitudinal smartphone sensing studies with the same protocol in diverse countries.

\begin{table*}
\centering
\caption{Terminology used in this study for training and testing approaches and target classes.}
\label{tab:keywords_explaination}
\resizebox{\textwidth}{!}{%
\begin{tabular}{>{\arraybackslash}m{3cm} >{\arraybackslash}m{13cm}}

\rowcolor[HTML]{EDEDED} 
\textbf{\makecell[l]{Terminology}}  & 
\textbf{Description} 
\\

Complex Daily Activity&
Based on prior studies that looked into complex activities of daily living \cite{saguna2013complex, rai2012mining, laput_sensing_2019}, we define these as activities that punctuate one's daily routine; that are complex in nature and occur over a non-instantaneous time window; and that have a semantic meaning and an intent, around which context-aware applications could be built. \\

\arrayrulecolor{Gray}
\hline 

Country-Specific &
This approach uses training and testing data from the same single country. 
Each country has its own model without leveraging data from other countries. As the name indicates, these models are specific to each country (e.g., a model trained in Italy and tested in Italy).\\

\arrayrulecolor{Gray}
\hline 

Country-Agnostic &
This approach assumes that data and models are agnostic to the country. Hence, a trained model can be deployed to any country regardless of the country of training. There are two types of country-agnostic phases:\newline (Phase I) This phase uses training data from one country and testing data from another country. This corresponds to the scenario where a trained machine learning model already exists, and we need to understand how it would generalize to a new country (e.g., a model trained in Italy and tested in Mongolia). \newline (Phase II) This phase uses training data from four countries and testing data from the remaining country. This corresponds to a scenario where the model was already trained with data from several countries, and we need to understand how it would generalize to a new country (e.g., a model trained with data from Italy, Denmark, UK, and Paraguay, and tested in Mongolia). \\

\arrayrulecolor{Gray}
\hline

Multi-Country&
This one-size-fits-all approach uses training data from all five countries and tests the learned model in all countries. This corresponds to the setting in which multi-country data is aggregated to build one single generalized model. However, this is also how models are typically built without considering aspects such as country-level diversity. \\

\arrayrulecolor{Gray2}
\hline 

\end{tabular}
}
\end{table*}

Our work uses a set of experimental approaches (country-specific, country-agnostic, and multi-country, described in Table~\ref{tab:keywords_explaination}), and model types (population-level and hybrid, described in Section~\ref{sec:inference}). With the support of rich multimodal smartphone sensing data collected in multiple countries under the same experimental protocol, we address three research questions:

\begin{itemize}[wide, labelwidth=!, labelindent=0pt]
    \item[\textbf{RQ1:}] How are complex daily activities expressed in different countries, and what smartphone sensing features are the most useful in discriminating different activities?
    
    \item[\textbf{RQ2:}] Is a generic multi-country approach well-suited for complex daily activity recognition? To which extent can country differences be accurately modeled by country-specific approaches?
    
    \item[\textbf{RQ3:}] Can complex daily activity recognition models be country-agnostic? In other words, how well do models trained in one or more countries generalize to unseen countries?
\end{itemize}{}

In addressing the above research questions, we provide the following contributions: 

\begin{itemize}[wide, labelwidth=!, labelindent=0pt]
    \item[\textbf{Contribution 1:}] We examined a novel smartphone sensor dataset and over 216K self-reports (including complex daily activities) collected from 637 college students in five countries (Denmark, Italy, Mongolia, Paraguay, and the United Kingdom) for over four weeks. To represent each activity self-report, we extracted around 100 features by processing multimodal smartphone sensor data (Table~\ref{tab:agg-features}). Moreover, we defined 12 complex daily activity classes based on participant responses, prevalence, and prior work. The list includes sleeping, studying, eating, watching something, online communication and social media use, attending classes, working, resting, reading, walking, sports, and shopping. On the one hand, we found that similar features are most informative for all countries for specific activities (e.g., sleep, shopping, walking). On the other hand, for some other activities, the most informative features vary across countries. Interestingly, however, they remain approximately similar across geographically closer countries. For example, the "sport" activity has the use of "health \& fitness apps" as a top feature across European countries. However, the feature was not prominent in Mongolia and Paraguay, where such physical activity-related app usage is lower. This divide is also visible in the “watching something” activity, which is influenced by the use of entertainment apps in European countries, and not in the other two countries. 
    
    \item[\textbf{Contribution 2:}] We defined and evaluated a 12-class complex daily activity inference task with country-specific, country-agnostic, and multi-country approaches (Table~\ref{tab:keywords_explaination}). We also used population-level (not personalized) and hybrid (partially personalized) models to evaluate how model personalization affects performance within and across countries. We show that the generic multi-country approach, which directly pools data from all countries (a typical approach in many studies), achieved an AUROC of 0.70 with hybrid models. Country-specific models perform the best for the five countries, with AUROC scores in the range of 0.79-0.89. These results suggest that even though multi-country models are trained with more data, models could not encapsulate all the information towards better performance, possibly due to the averaging effect of diverse behaviors across countries. The country-specific approach consistently worked better. 
    
    \item[\textbf{Contribution 3:}] With the country-agnostic approach, we found that models do not generalize well to other countries, with all AUROCs being below 0.7 in the population-level setting. With hybrid models, personalization increased the generalization of models reaching AUROC scores above 0.8, but not up to the same level as country-specific hybrid models. Moreover, even after partial personalization, we observed that models trained in European countries performed better when deployed in other European countries than in Mongolia or Paraguay. This shows that in addition to country diversity, behavior and technology usage habits could be what mediates the performance of models in different countries. In light of these findings, we believe that human-computer interaction and ubiquitous computing researchers should be aware of machine learning models' geographic sensitivities when training, testing, and deploying systems to understand real-life human behavior and complex daily activities. We also highlight the need for more work to address the domain shift challenge in multimodal mobile sensing datasets across countries.
\end{itemize}

To the best of our knowledge, this is the first study that focuses on the use of multimodal smartphone sensing data for complex daily activity recognition, while examining the effect of country-level diversity of data on complex activity recognition models with a large-scale multi-country dataset, and highlighting domain shift-related issues in daily activity recognition, even when the same experimental protocols are used to collect data in different countries.

The paper is organized as follows. In Section~\ref{sec:relatedwork}, we describe the related work and background. Then, we describe the dataset in Section~\ref{sec:dataset}. In Section~\ref{sec:descriptive}, we present the descriptive and statistical analysis regarding important features. We define and evaluate inference tasks in Section~\ref{sec:inference} and Section~\ref{sec:results}. Finally, we end the paper with the Discussion in Section~\ref{sec:discussion} and the Conclusion in Section~\ref{sec:conclusion}.
\section{Background and Related Work}\label{sec:relatedwork}

\subsection{Mobile Sensing}

In prior work, researchers have collected and analyzed mobile sensing data to understand various attributes of a particular population. Depending on the study, that goal can be put under coarse categories such as behavior, context, and person-aspect recognition \cite{meegahapola2020smartphone}. Behavior recognition is aimed at understanding user activities broadly. Person aspect recognition looks into understanding demographic attributes (e.g., sex, age, etc.), psychology-related attributes (e.g., mood, stress, depression, etc.), and personality. Finally, context recognition identifies different contexts (e.g., social context, location, environmental factors, etc.) in which mobile users operate. 

Regarding behavior recognition, there are studies that aimed to capture binary (sometimes three) states of a single complex activity/behavior such as eating (e.g., eating meals vs. snacks \cite{biel_bitesnbits_2018}, overeating vs. undereating vs. as usual eating \cite{meegahapola_one_2021}), smoking (e.g., smoking or not \cite{mcclernon2013your}) or drinking alcohol (e.g., drinking level \cite{phan2020understanding, bae2017detecting}, drinking or not \cite{santani_drinksense_2018}). Another study used the action logs of an audio-based navigation app to predict its usage and understand what drives user engagement \cite{low_vision_app_engagement}. Then, regarding person aspects, the MoodScope system \cite{likamwa_moodscope_2013} inferred the mood of smartphone users with a multi-linear regression based on interactions with email, phone, and SMS, as well as phone location and app usage. Servia-Rodriguez et al. \cite{servia-rodriguez_mobile_2017} observed a correlation between participants' routines and some psychological variables. They trained a deep neural network that could predict participants' moods using smartphone sensor data. Additionally, Khwaja et al. \cite{khwaja_modeling_2019} developed personality models based on random forests using smartphone sensor data. Finally, context recognition is aimed at detecting the context around behaviors and activities. \cite{meegahapola_alone_2020} used sensing data from Switzerland and Mexico to understand its relation to the social context of college students when performing eating activities. More specifically, they built an inference model to detect whether a participant eats alone or with others. Similarly, \cite{meegahapola_examining_2021} examined smartphone data from young adults to infer the social context of drinking episodes using features from modalities such as the accelerometer, app usage, location, Bluetooth, and proximity. In this case, context detection is two-fold: it's based on the number of people in a group, and on their relationship to the participant (e.g., alone, with another person, with friends, with colleagues). Similarly, mobile sensing studies attempted to infer other contexts, psychological traits, and activities by taking behavior and contexts sensed using smartphone sensors as proxies \cite{meegahapola2020smartphone, cornet2018systematic, hoseini2013survey}. 

One common aspect regarding most of these studies is that they were done in the wild, focused on two or three-class state inference, and sensing is not fine-grained (i.e., using behavior and context as proxies to the dependent variable). This paper follows a similar approach with a dataset captured real life, using multimodal smartphone sensor data, and taking behavior and context as proxies for our dependent variable. However, in this study, the target attribute entails a 12-class daily activity recognition problem that is complex and novel compared to prior work. In addition, we are interested in examining model performance within and across five countries, with and without partial personalization. 

\subsection{Activity Recognition}

Human activity recognition (HAR) aims to understand what people are doing at a given time. Large-scale datasets issued from the activity of smartphone users have a lot of potential in solving that task. This ``digital footprint" has been used to re-identify individuals using credit-card metadata \cite{credit_card_reidentification}: it has been shown that only 4 data points are required to re-identify 90\% of individuals. While the same approach could be followed using smartphone sensing data, our main focus is activity recognition at a single point in time rather than using time series for re-identification. We will focus on two types of activity recognition techniques: wearable-based and smartphone-based \cite{straczkiewicz2021systematic}.

\subsubsection{Wearable-based HAR} 

In wearable-based activity recognition, the users wear sensors such as wearable accelerometers from which the data is analyzed and classified to detect activities. For example, in healthcare, wearable-based HAR can be used to analyze gait and prevent falling or monitor physical activity and observe health outcomes \cite{liu_overview_2021}. The wearable-sensing trend emerged two decades ago and relied on custom-designed wearable sensors \cite{farringdon_wearable_1999, park_wearable_1999}, which were backed by encouraging findings in health research. With time, custom sensors were replaced by commercial fitness or activity trackers. Unfortunately, applying these findings to real-world settings was rare due to the high cost of producing custom sensors, the difficulty distributing devices to a broad audience, and their unpopularity among some users \cite{coorevits_rise_2016}. This restricted most studies using wearables to performing experiments in a controlled environment or in the wild with smaller populations. However, wearable-based HAR models that could recognize simple activities are currently deployed across many commercial wearable devices.

\subsubsection{Smartphone-based HAR}

With the popularity of smartphones in the past two decades, the problems of wearable-based HAR were solved. Reality Mining \cite{eagle_reality_2006} is a pioneering study in the field of mobile sensing: it showed the utility of mobile sensing data in a free-living setting. In smartphone-based activity recognition, people do not need to use wearable sensors. Instead, the system relies on a smartphone that is always on and stays closer to its user. Smartphones replace wearable devices as the former contains multiple sensors such as an accelerometer, gyroscope, GPS, proximity, or thermometer. Nevertheless, smartphones capture data at multiple positions (e.g., a pocket, hand, or handbag), which introduces a bias in sensor measurements as they are position-dependent \cite{yang_pacp_2016}.

Regardless of the device used, most prior activity recognition tasks have been done in lab-based/controlled settings where accurate ground truth capture is possible \cite{straczkiewicz_systematic_2021}. The prime goal of such studies is to increase the accuracy of activity recognition models with precise ground truth and sensor data collection (e.g., by placing sensors on fixed body positions, recording ground truth with videos, etc.). However, these studies are hard to scale and do not capture the real behavior of participants, and this is especially true for complex daily activities \cite{saguna2013complex}. For example, a person's behavior when studying, working, or shopping in an unconstrained environment can not be replicated in a lab. On the other hand, some studies are done in the wild \cite{laput2019sensing, saguna2013complex}, where the ground truth and sensor data collection might not be that precise but allow capturing complex daily activities in a naturalistic setting. Our study is similar, where our intention was to take a more exploratory stance, build country-level diversity-aware models, and compare their performance within and across different countries.  

\subsection{Activity Types}

One crucial difference across existing studies is in the selection of activities. A majority of studies work towards the recognition of simple activities. For example, Straczkiewicz et al. \cite{straczkiewicz_systematic_2021} classified activities into groups such as posture (lying, sitting, standing), mobility (walking, stair claiming, running, cycling), and locomotion (motorized activities). Laput and Harrison \cite{laput_sensing_2019} called such activities coarse or whole-body. Activities belonging to these groups are directly measurable from one or more proxies (e.g., inertial sensor unit, location). For example, when considering the accelerometer, each activity has a distinct pattern on the different axes \cite{dernbach_simple_2012}. However, they constitute a small subset of activities performed by people in daily lives \cite{saguna2013complex, rai2012mining, cobian2013active}.

Notice that some of the simple activities described above are
usually part of more complex activities (e.g., sitting while eating, walking while shopping). Dernbach et al. \cite{dernbach_simple_2012} defined complex activities as a series of multiple actions, often overlapping. Along with Bao et al. \cite{bao_activity_2004}, they used the same techniques to recognize both simple and complex activities. This results in weaker performances for complex activities since their structure is more complicated. Another approach is considering complex activities hierarchically by using combinations of simple activities to predict more complex ones. Huynh et al. \cite{huynh_discovery_2008} characterized user routines as a probabilistic combination of simple activities. Blanke et al. \cite{choudhury_daily_2009} used a top-down method that first identifies simple activities to recognize complex ones. However, this requires pre-defining simple activities and mappings to complex activities. Some studies focus on detecting binary episodes of a single complex activity or a specific action. For example, the Bites'n'Bits study \cite{biel_bitesnbits_2018} examined the contextual differences between eating a meal and a snack, and presented a classifier able to discriminate eating episodes among students. Likewise, DrinkSense \cite{santani_drinksense_2018} aimed at detecting alcohol consumption events among young adults on weekend nights. Unfortunately, such task-specific classifiers will perform poorly when exposed to situations they were not trained on. 

In this study, we focus on a majority of complex daily activities (11 out of 12 and one simple activity: walking) derived by considering over 216K self-reports from college students in five countries. In this context, drawing from prior studies that looked into activities of daily living \cite{saguna2013complex, rai2012mining}, for the scope of this paper, we define complex activities as \emph{"activities that punctuate one's daily routine; that are complex in nature and occur over a non-instantaneous time window; and that have a semantic meaning and an intent, around which context-aware applications could be built"}. While it is impossible to create a classifier that could recognize all complex human activities, we believe the classifier we propose captures a wide range of prevalent activities/behaviors, especially among young adults. 

\subsection{Diversity-Awareness in Smartphone Sensing}

Research in the field of smartphone sensing, including the studies mentioned above, lacks diversity in their study populations \cite{meegahapola2020smartphone}. Regarding country diversity, with a few exceptions \cite{servia-rodriguez_mobile_2017, khwaja_modeling_2019}, most experiments were conducted in a single country or rarely two. This can be problematic with respect to the generalization of findings since smartphone usage differs across geographic regions, which can lead to different patterns being observed in, for example, two populations of different genders or age range \cite{del_rosario_comparison_2014}. Khwaja et al. stressed the importance of diversity awareness in mobile sensing \cite{khwaja_modeling_2019}. Moreover, experiments performed in a controlled setting usually can not accommodate many participants. While this makes the whole process lighter and more manageable, it also restricts the generalization of results to a broader free-living audience \cite{sasaki_performance_2016, van_hees_impact_2013}. According to Phan et al. \cite{phan_mobile_2022}, cross-country generalizability is the extent to which findings apply to different groups other than those under investigation. 

Diversity awareness and model generalization are two essential aspects, as they will allow an activity recognition system to be deployed and to perform well across different user groups and countries \cite{meegahapola2023generalization, Schelenz2021The}. In computer vision research, the lack of diversity has been repeatedly shown for specific attributes such as gender and skin color \cite{raji_actionable_2019, cuthbertson_self-driving_2019, karkkainen2021fairface}. In natural language processing and speech research, not accounting for dialects in different countries could marginalize groups of people from certain countries~\cite{prabhakaran2022cultural}. Hence, ignoring country diversity when developing AI systems could harm users in the long run by marginalizing certain groups of people \cite{prabhakaran2022cultural}. In this context, smartphone sensing studies that consider country-level diversity are still scarce \cite{phan_mobile_2022}. This could be due to the lack of large-scale datasets, logistical difficulties in data collection in different countries, and studies being time and resource-consuming. Khwaja et al. \cite{khwaja_modeling_2019} built personality inference models using smartphone sensor data from five countries and showed that such models perform well when tested in new countries. To the best of our knowledge, their study is one of the first to investigate the generalization of smartphone sensing-based inference models across different countries. In our work, we focus on complex daily activity recognition with smartphone sensing and aim to uncover and examine model behavior in multi-country settings. 

\subsection{Human-Centered Aspects in Smartphone Usage}

Our literature review has so far focused on the technical aspects such as data collection or target variables. We now discuss the impact of smartphone usage on individuals and society, which is studied by various disciplines in the social sciences. Previous work includes the study of smartphone dependence among young adults, where it was found that problematic smartphone use varies by country and gender \cite{self_reported_phone_dependence, addictive_behaviour_modelling}, and those specific activities such as social networking, video games, and online shopping contribute to the addiction \cite{self_reported_phone_dependence, smartphone_addiction_24_countries}. Another study \cite{Rathod_Ingole_Gaidhane_Choudhari_2022} summarized findings on correlations between smartphone usage and psychological morbidities among teens and young adults. Excessive smartphone usage could lead to emotional difficulties, impulsivity, shyness, low self-esteem, and some medical issues such as insomnia, anxiety, or depression. From a sociological standpoint, Henriksen et al. \cite{soc10040078} studied how smartphones impact interactions in cafés and defined three concepts of social smartphone practices. \emph{Interaction suspension} (e.g., your friend goes to the bathroom), which can lead to using the smartphone to appear occupied or to avoid uncomfortable situations while being alone. \emph{Deliberate interaction shielding} corresponds to situations where one suspends an ongoing interaction to answer a phone call or a text message, whether it is an emergency or just in fear of missing out. \emph{Accessing shareables}, which leads to a collective focus on shared content (e.g., pictures or short videos), giving the smartphone a role of enhancing face-to-face social interactions rather than obstructing them. Nelson and Pieper \cite{Nelson_Pieper_2022} showed that smartphone attachment ``inadvertently exacerbates feelings of despair while simultaneously promises to resolve them", thus trapping users in negative cycles.

According to Van Deursen et al. \cite{addictive_behaviour_modelling}, older populations are less likely to develop addictive smartphone behaviors. While they are often associated with younger generations, smartphones are slowly gaining popularity among older generations as they are coming up with creative ways to integrate them into their habits. Miller et al. \cite{global_smartphone} investigated the role that smartphones play in different communities across nine countries. Through 16-month-long ethnographies, they showed that various groups of people have specific ways of taking ownership of their smartphones through apps, customization, and communication. For example, in Ireland, smartphones are used by the elderly in many of their daily activities, and in Brazil, the usage of messaging applications for health have lead to the creation of a manual of best practices for health through such applications. More globally, smartphones can help users stay in touch with their extended families or distant friends, a feature that has been particularly important during the 2020 global pandemic. In this paper, we attempt to uncover country-specific smartphone usage patterns through multimodal sensing data. While these insights may not have the depth that field observations provide, they represent a starting point for future research to draw upon.

\begin{figure*}
    \includegraphics[width=0.7\textwidth]{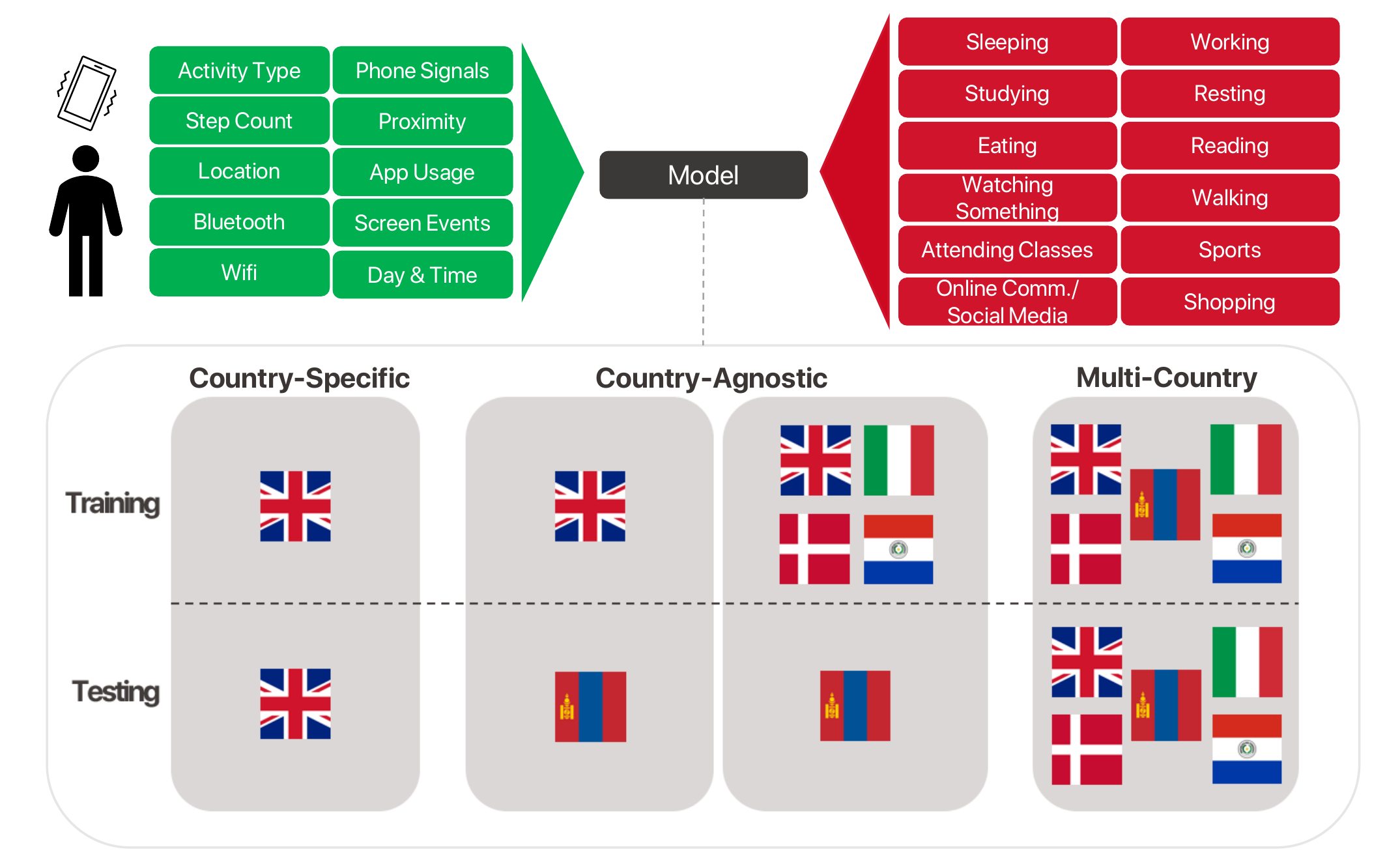}
    \caption{High-level overview of the study. The study uses continuous and interaction sensing modalities and different approaches (country-specific, country-agnostic, and multi-country) to infer complex daily activities.}
    \label{fig:overview}
    \Description{High-level overview of the study is given in this diagram. The study uses continuous and interaction sensing modalities and different approaches such as country-specific, country-agnostic, and multi-country, to infer complex daily activities.}
\end{figure*}

Hence, all while considering these factors, we aim to examine smartphone sensing-based inference models for complex daily activity recognition with country-specific, country-agnostic, and multi-country approaches, as described in Figure~\ref{fig:overview}.

\section{Data, Features, and Target Classes}\label{sec:dataset}

\subsection{Dataset Information}

To address our research questions, we collected a smartphone sensing dataset regarding the everyday behavior of college students for four weeks during November 2020, in the context of the European project "WeNet: The Internet of Us" \footnote{The dataset is planned to be released for research purposes after the end of the project, by complying with all regulations governing the data collection protocol within and outside the European Union. Hence, future plans for dataset access will be made available on the project website: \url{https://www.internetofus.eu/}}. The study procedure is summarized in a technical report \cite{giunchiglia2022aworldwide}. The sample consisted of both undergraduate and graduate students. This dataset was collected to study the effect of the diversity of study participants on social interactions and smartphone sensor data. The dataset contains over 216K self-reported activities collected from 637 college students living in five countries (ordered by the number of participants): Italy, Mongolia, the United Kingdom, Denmark, and Paraguay. All data were collected using Android smartphones with the same mobile app. Table \ref{tab:num-participants} shows the distribution of participants across countries. Moreover, the data were collected with a protocol compliant with the EU General Data Protection Regulation (GDPR) and with each non-EU country's rules. In addition, written approvals from the ethical review boards (or similar entities) were acquired by each participating university, separately.

\begin{table*}
    \centering
    \caption{A summary of participants of the data collection. Countries are sorted based on the number of participants.}
    \label{tab:num-participants}
    \begin{tabular}{llcccr}
        \rowcolor[HTML]{EDEDED} 
        \textbf{University} &
        \textbf{Country} &
        \textbf{Participants} &
        \textbf{$\mu$ Age ($\sigma$)} &
        \textbf{\% Women} &
        \textbf{\# of Self-Reports}
        \\

\arrayrulecolor{Gray}

        University of Trento &
        Italy &
        259 &
        24.1 (3.3) &
        58 &
        116,170 
        \\

\arrayrulecolor{Gray}
\hline 
        
        National University of Mongolia &
        Mongolia &
        224 &
        22.0 (3.1) &
        65 &
        65,387 
        \\

\arrayrulecolor{Gray}
\hline 
        
        \makecell[l]{London School of Economics\\ \& Political Science} &
        UK &
        86 &
        26.6 (5.0) & 
        66 &
        20,238 
        \\

\arrayrulecolor{Gray}
\hline 
        
        \makecell[l]{Universidad Católica \\"Nuestra Señora de la Asunción"} &
        Paraguay &
        42 &
        25.3 (5.1) &
        60 &
        6,998 
        \\

\arrayrulecolor{Gray}
\hline

        Aalborg University &
        Denmark  &
        26 &
        30.2 (6.3) & 
        58 &
        7,461
        \\

\arrayrulecolor{Gray2}

\rowcolor[HTML]{EDEDED} 
        Total/Mean &
        &
        637 &
        24.0 (4.3) &
        62 &
        216,254
        \\

    \end{tabular}
\end{table*}

The first phase of the data collection obtained questionnaire data about the participants, their habits, social relations, individual practices (e.g., physical activities, leisure), and skills (personal and interpersonal). This data was aimed at capturing different aspects of diversity, including observable characteristics such as demographics as well as less observable aspects such as personality, traits, skills, values, and relations \cite{schelenz_theory_2021}. The second phase collected data through a smartphone application. Participants filled out time diaries multiple times throughout the day. Participants were asked about their sleep quality and expectations at the start of the day. At the end of the day, they had to report how their day went. At every hour, they had to self-report what they were doing (current activity, using a drop-down list of 34 activities), location (a list of 26 semantic locations), social context (a list of 8 social contexts), and mood (valence was captured similar to \cite{likamwa_moodscope_2013} with a five-point scale). The app continuously collected data from more than thirty smartphone sensors, which can be broken down into two categories \cite{meegahapola2020smartphone}: continuous sensing modalities such as the simple activity type (derived using inertial sensors and location with the Google Activity Recognition API \cite{GoogleActivity2022}), step count, location, WiFi, Bluetooth, phone signal, battery, and proximity; and interaction sensing modalities such as application usage time, screen usage episode counts and time, notification clicking behaviors, and user presence time. 

\subsection{Deriving Features}

The choice of the dataset's format is key for the rest of the study. A \emph{tabular} dataset centered around activities or events enables the handcrafting of a multitude of sensor-specific features discussed in prior literature \cite{santani_drinksense_2018, meegahapola2020smartphone, canzian2015trajectories, servia-rodriguez_mobile_2017, meegahapola_one_2021}. This later enables the use of traditional machine learning methods. However, a {temporal} dataset relies mainly on raw sensor measurements in the form of time series (i.e., raw accelerometer and gyroscope data in typical activity recognition). This approach allows deep learning methods to extract and learn relevant high-level features automatically. Past research \cite{zhu_efficient_2019, aviles-cruz_coarse-fine_2019} has shown that using deep learning techniques yields better-performing HAR classifiers. These studies typically include simple activities that are easier to detect with inertial sensors than the more complex ones we are interested in. This is particularly important in remote study/work settings, where many activities are performed while at home. Therefore, we chose to perform the analysis using a tabular dataset with the heterogeneous handcrafted features described below.

We aggregated all sensor measurements with self-reports to create features using smartphone data. We followed a time-window-based approach similar to prior studies on event-level inferences \cite{meegahapola2020smartphone, straczkiewicz2021systematic, servia-rodriguez_mobile_2017}. Hence, we used 10 minutes before and after each self-report and aggregated sensor data in the corresponding 20-minute interval \footnote{We conducted experiments with different time windows between 5 minutes and 25 minutes. We did not go beyond 25 minutes because it would lead to overlapping sensor data segments, hence leaking data between data points. 20-minute window performed the best out of the examined time windows. For brevity, we only present results with the 20-minute window. Shorter windows might not have performed reasonably because they do not capture enough contextual information to make the inference. Prior work too has shown that large time windows might be suitable to detect binary activities \cite{bae_detecting_2017, meegahapola_examining_2021, biel_bitesnbits_2018}}. While traditional and inertial sensor-based recognition of simple activities attempts to capture repetitive moments using deep learning with a smaller time window, that method is not applicable here because we attempt to capture a set of non-repetitive activities that last longer. In addition, we consider behavior and context sensed with the smartphone as a proxy to the target activity, similar to prior ubicomp studies \cite{mcclernon2013your, meegahapola_one_2021, servia-rodriguez_mobile_2017}. So, the corresponding features generated using sensing modalities are shown in Table \ref{tab:agg-features}. More details on how each sensing modality was pre-processed can be found in \cite{meegahapola2023generalization}. In addition to sensor data features, we added a feature that describes the time period of the day when the activity occurred and a weekend indicator. While there is no agreement as to how a day could be split into morning, afternoon, evening, and night in the literature \cite{obuchi2020predicting, wang2020predicting, wang2017predicting, nepal2020detecting, wang2022first}, we defined five time periods: morning from 6 AM to 10 AM, noon from 10 AM to 2 PM, afternoon from 2 PM to 6 PM, evening from 6 PM to 10 PM, and night from 10 PM to 6 AM, and included it as another feature that could be used in training machine learning models. 

\begin{table*}
    \caption{Summary of 108 features extracted from raw sensing data, aggregated around activity self-reports using a time window.}
    \label{tab:agg-features}
    \centering
    \resizebox{0.9\textwidth}{!}{%
    \begin{tabular}{{p{0.2\linewidth}p{0.8\linewidth}}}
        \rowcolor[HTML]{EDEDED} 
        \textbf{Modality} & \textbf{Corresponding Features and Description} \\

        Location & radius of gyration, distance traveled, mean altitude calculated similarly to prior work \cite{canzian2015trajectories} \\
        
        \arrayrulecolor{Gray}
        \hline 
        Bluetooth {[}LE, normal{]} & number of devices (the total number of unique devices found), mean/std/min/max RSSI (Received Signal Strength Indication -- measure close/distant closer devices are) \cite{santani_drinksense_2018} \\
        
        \arrayrulecolor{Gray}
        \hline 
        WiFi & connected to a network indicator, number of devices (the total number of unique devices found), mean/std/min/max RSSI \cite{santani_drinksense_2018} \\
        
        \arrayrulecolor{Gray}
        \hline  
        Cellular {[}GSM, WCDMA, LTE{]} & number of devices (the total number of unique devices found), mean/std/min/max phone signal strength \cite{santani_drinksense_2018} \\
        
        \arrayrulecolor{Gray}
        \hline 
        Notifications & notifications posted (the number of notifications that came to the phone), notifications removed (the number of notifications that were removed by the participant) -- these features were calculated with and without duplicates. \cite{likamwa_moodscope_2013}\\
        
        \arrayrulecolor{Gray}
        \hline  
        Proximity & mean/std/min/max of proximity values \cite{bae_detecting_2017} \\
        
        \arrayrulecolor{Gray}
        \hline 
        Activity & time spent doing the following simple activities: still, in\_vehicle, on\_bicycle, on\_foot, running, tilting, walking, other (derived using the Google Activity Recognition API \cite{GoogleActivity2022}) \\
        
        \arrayrulecolor{Gray}
        \hline 
        Steps & steps counter (steps derived using the total steps since the last phone turned on), steps detected (steps derived using event triggered for each new step) \cite{das2022semantic} \\
        
        \arrayrulecolor{Gray}
        \hline 
        Screen events & touch events (number of phone touch events), user presence time (derived using android API that indicate whether a person is using the phone or not), number of episodes (episode is from turning the screen of the phone on until the screen is turned off), mean/min/max/std episode time (a time window could have multiple episodes), total time (total screen on time within the time window) \cite{likamwa_moodscope_2013, bae_detecting_2017, meegahapola_one_2021} \\
        
        \arrayrulecolor{Gray}
        \hline 
        App events & time spent on apps of each category derived from Google Play Store \cite{likamwa_moodscope_2013, santani_drinksense_2018}: action, adventure, arcade, art \& design, auto \& vehicles, beauty, board, books \&
        reference, business, card, casino, casual, comics, communication,
        dating, education, entertainment, finance, food \& drink, health \&
        fitness, house, lifestyle, maps \& navigation, medical, music, news \&
        magazines, parenting, personalization, photography, productivity,
        puzzle, racing, role-playing, shopping, simulation, social, sports,
        strategy, tools, travel, trivia, video players \& editors, weather, word   \\
        
        \arrayrulecolor{Gray}
        \hline 
        Time \& Day & hour of the day, period of the day (morning, noon, afternoon, evening, night), weekend indicator (weekday or weekend) \cite{biel_bitesnbits_2018, meegahapola_one_2021} \\
        
        \arrayrulecolor{Gray2}
        \hline 
    \end{tabular}}
\end{table*}

\subsection{Determining Target Classes}

\begin{figure*}
    \includegraphics[width=0.8\textwidth]{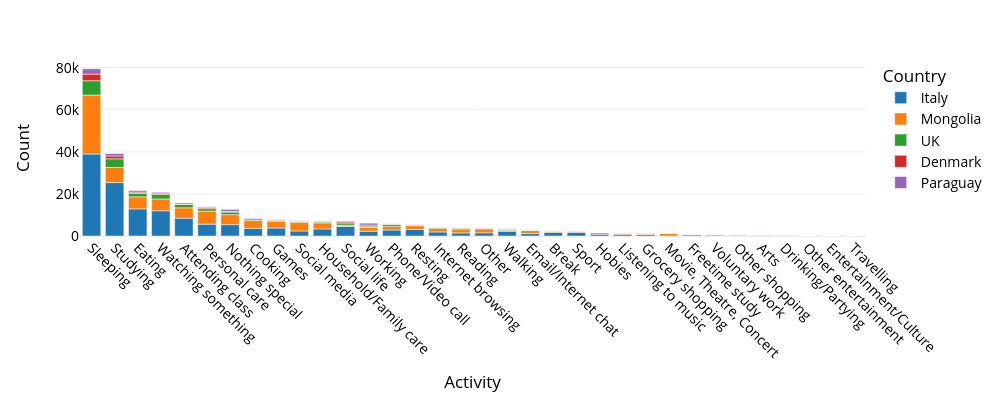}
    \caption{The original distribution of target classes before any filtering or merging was done.}
    \Description{The original distribution of target classes before any filtering or merging was done. X-axis shows the activities. Y-axis shows the country, separated by countries. Activities with the highest number of self-reports include sleeping, studying, eating, watching something, and attending class.}
    \label{fig:target-dist-before}
\end{figure*}

\begin{figure}
    \includegraphics[width=0.5\textwidth]{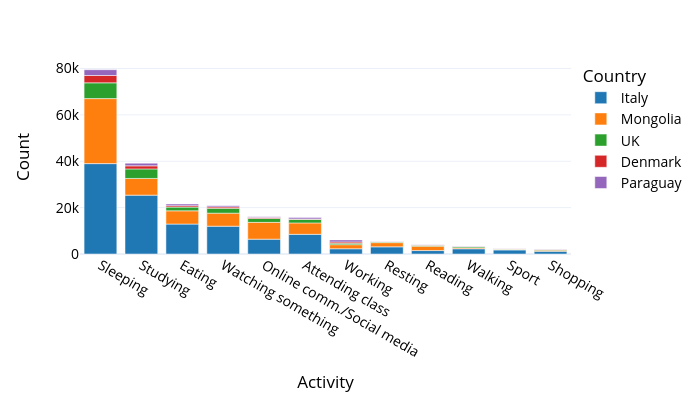}
    \caption{Distribution of target classes after removing classes that are semantically broad or lack data.}
    \Description{Distribution of target classes after removing classes that are semantically broad or lack data. Now the 12 classes include sleeping, studying, eating, watching something, online communication or social media, attending class, working, resting, reading, walking, sport, and shopping.}
    \label{fig:target-dist-after}
\end{figure}

Hourly self-reports required participants to log what they were doing at the time by selecting an activity from a predefined list of thirty-four items. These items were derived based on prior work \cite{giunchiglia2017personal, zhang2021putting}. By looking at their distribution in different countries (Figure \ref{fig:target-dist-before}), one can quickly notice that they are highly unbalanced. The remote work/study constraints during the time of data collection were one of the causes behind this imbalance, because activities such as traveling, walking, or shopping would have been more popular if mobility was not restricted. A closer look at the list of activities shows that some classes are too broad in terms of semantic meaning. Hence, similar to prior work that narrowed down activity lists based on various aspects \cite{laput_sensing_2019}, we narrowed down the original list of activities into 12 categories to capture complex daily activities that are common enough in the daily lives of people, especially in a remote work/study setting. For example, under ``hobbies'', one can be playing the piano or painting, and the two do not entail the same smartphone usage and are not common enough. Similarly, “social life” is too broad, as one could be in a bar, a restaurant, or a park. Moreover, to mitigate the class imbalance problem, we decided to filter the target classes. First, classes that had similar semantic meanings were merged: this is the case of eating and cooking, and social media and internet chatting. Classes representing a broad activity were removed, such as personal care, household care, games, and hobbies. Finally, classes that did not have enough data in all countries were removed, such as listening to music, movie, theatre, concert, and free-time study. Filler classes such as ``nothing special'' or ``other'' were also removed. This filtering reduced the number of target classes to twelve, and their updated distribution is shown in Figure \ref{fig:target-dist-after}. These classes entail activities performed during daily life that are complex in nature and have a semantic meaning around which context-aware applications could be built. Moreover, the selected activities also align with prior work that looked into complex daily activity recognition \cite{saguna2013complex}.

\section{How are activities expressed in different countries, and what smartphone features are most discriminant? (RQ1)}\label{sec:descriptive}

\begin{figure*}
    \includegraphics[width=\textwidth]{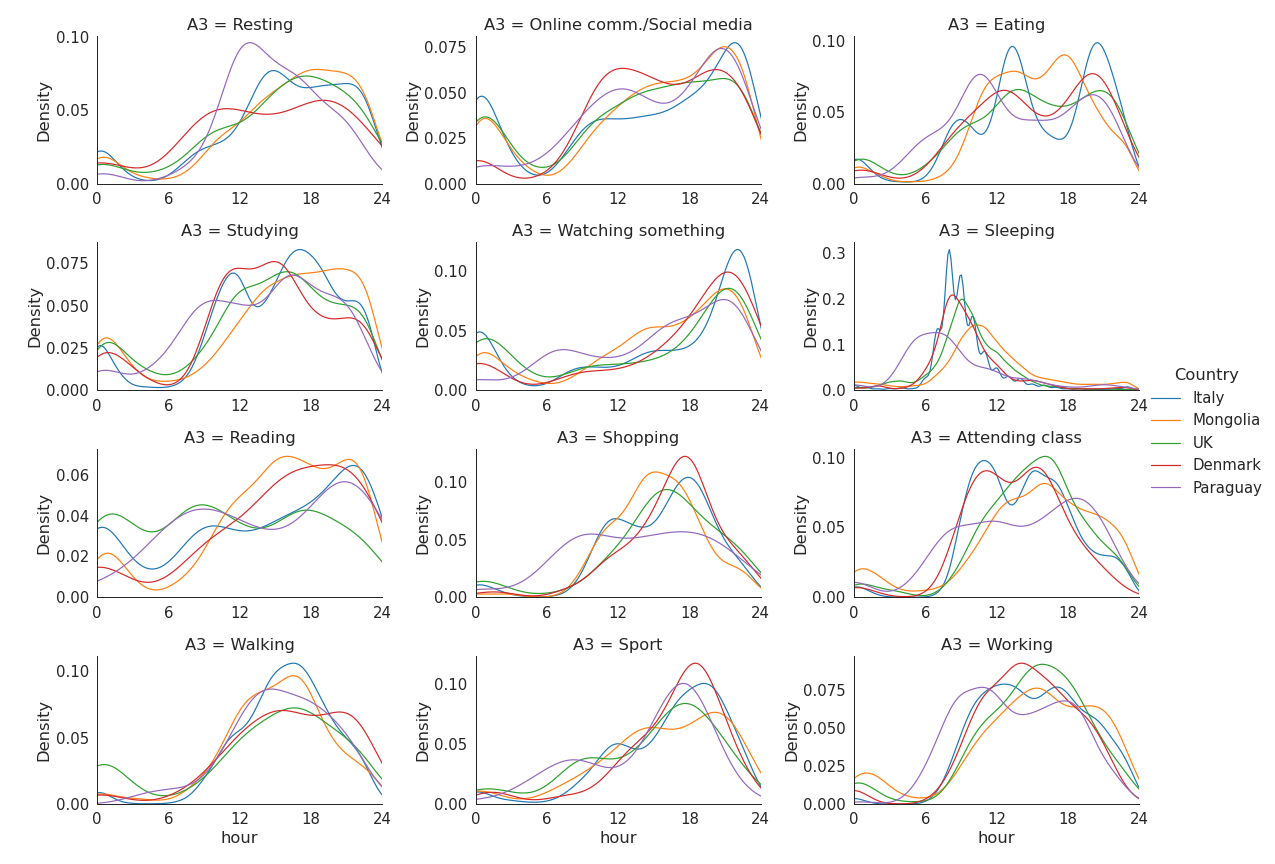}
    \caption{Density functions of target classes as a function of the hour of day in each country.}
    \Description{Density functions of target classes as a function of the hour of the day in each country. In each diagram, the x-axis refers to the hour of the day, and the y-axis refers to the density of each activity.}
    \label{fig:activities-per-hour}
\end{figure*}

To understand the distribution of activities in each country and to determine the influence of features on the target, we provide a descriptive and statistical analysis of the dataset in this section, hence shedding light on \textbf{RQ1}.

\subsection{Hourly Distribution of Activities}

The activities we consider all seem to occur at different times: people tend to sleep at night, work during the day, and eat around noon and in the evening. However, not all schedules are the same, especially not across different countries \cite{fisher_daily_2010, esteban_ortiz-ospina_time_2020}. We reported the density function of each target class at different hours of the day in Figure \ref{fig:activities-per-hour}. In each diagram, the x-axis refers to the hour of the day, and the y-axis refers to the density of each activity. On an important note, while most activities were reported as they were being performed, in the case of sleeping, participants reported the activity {after} they woke up and still in bed, meaning that peaks for that activity could also be interpreted as ``waking up''. This was later confirmed with many participants in all countries during post-study interviews. This also makes the time of the day less informative when inferring the sleeping activity.

A first look at the distribution shows the ``expected'' patterns, such as a peek of sleeping during the night or peaks around eating times for lunch and dinner. Notice that participants from Paraguay tend to sleep less than others, reflecting that they start working and resting earlier in the day. Online communication and social media usage happen around noon, coinciding with a break from classes and lunchtime, followed by a high peak towards the end of the day. This is in line with prior studies that showed that depending on the location context and hour of the day, the use of certain social media applications (i.e., Twitter) could differ \cite{deng2018social}. Moreover, we also observe country differences in hourly social media and online communication app usage patterns as reported by users. For example, between noon and 6 pm, there is a dip in the usage of these types of apps in Italy, Paraguay, and Denmark, whereas that pattern is not visible in the UK. Prior work has also studied social media app usage and adoption-related differences, especially across countries. As per those studies, such usage differences could result from cultural characteristics within countries and from motives of people for using different apps \cite{alsaleh2019cross, lim2014investigating}. Most leisure activities (reading, shopping, sport, watching something) happen towards the end of the day, right when students have finished their classes. 

Another activity that showed clear cross-country differences is ``Eating''. We can observe that Italians tend to eat later than others, which hints at their Mediterranean customs \cite{tobin_what_2018}. Italy also showed two clear peaks for lunch and dinner with a sharp dip in between the two meals. The dip is less visible in other countries, indicating that meals are more spread out across different times. Moreover, the dinner peaks for all countries except Mongolia were peaking on or after 6 pm, whereas in Mongolia, it was before 6 pm. These findings suggest that the hour of the day could indicate whether people are eating or not---slightly differently in Italy, Mongolia, and other countries. In fact, prior studies that used mobile sensors for studies regarding eating behavior showed that the hour of the day is an important feature in predicting aspects related to eating \cite{biel_bitesnbits_2018, meegahapola_one_2021}. To add to that, prior studies have also pointed out that meal times, frequency, and sizes too could differ between countries \cite{chiva1997cultural}, even within Europe \cite{southerton2012behavioural}. Finally, the activity ``walking'' had more or less similar distributions across countries. In fact, a smartphone-based activity tracking study by Althoff et al. \cite{althoff2017large} mentioned that the average number of steps walked by people across Italy, the UK, Denmark, and Mongolia were in the same ballpark (i.e., around 5000-6000 daily steps).

\subsection{Statistical Analysis of Features}

To understand the importance of each smartphone sensing feature in discriminating each target activity from others, we reported in Table \ref{tab:statistical-analysis} the top three features and their ANOVA (Analysis of variance) F-values \cite{kim2014analysis} for each activity and each country. The goal is to identify features that define an activity and how those differ across countries. We consider each country-activity pair alone to find features that influence the classification task in a binary setting (i.e., determining whether the participant is sleeping or not, studying or not, eating or not, etc.).

\begin{table*}
    \caption{ANOVA F-values (F) with p-value $< 0.05$ for each target activity
    and each country. The best feature is the first in the list. Comparing
    F-values are only valid locally within the same activity and country.}
    \label{tab:statistical-analysis}
    \resizebox{1\textwidth}{!}{%
    \begin{tabular}{cllllllllll}
    
    \rowcolor[HTML]{EDEDED} 
    \multicolumn{1}{l}{\cellcolor[HTML]{EDEDED}} & \multicolumn{2}{c}{\cellcolor[HTML]{EDEDED}\textbf{Italy}} & \multicolumn{2}{c}{\cellcolor[HTML]{EDEDED}\textbf{Mongolia}} & \multicolumn{2}{c}{\cellcolor[HTML]{EDEDED}\textbf{UK}} & \multicolumn{2}{c}{\cellcolor[HTML]{EDEDED}\textbf{Denmark}} & \multicolumn{2}{c}{\cellcolor[HTML]{EDEDED}\textbf{Paraguay}} 
    \\ 
    
    \multicolumn{1}{l}{} & \textit{Feature} & \textit{F} & \textit{Feature} & \textit{F} & \textit{Feature} & \textit{F} & \textit{Feature} & \textit{F} & \textit{Feature} & \textit{F} 
    \\
    
    \arrayrulecolor{Gray}
        \hline 
    
     & app\_tools & 5423 & day\_period & 6623 & day\_period & 1632 & day\_period & 354 & app\_not-found & 510 
     \\
     
     & day\_period & 4439 & app\_not-found & 3595 & screen\_max\_episode & 603 & screen\_max\_episode & 249 & noti\_removed\_wo\_dups & 348 \\
     
    \multirow{-3}{*}{\textbf{Sleeping}} & screen\_max\_episode & 2498 & noti\_removed\_wo\_dups & 1052 & screen\_time\_per\_episode & 534 & screen\_time\_per\_episode & 156 & notifications\_posted\_wo\_dups & 289 
    \\
    
    \rowcolor[HTML]{EFEFEF} 
    \cellcolor[HTML]{EFEFEF} & screen\_max\_episode & 1447 & day\_period & 683 & screen\_time\_total & 446 & screen\_max\_episode & 241 & app\_video players \& editors & 147 
    \\
    
    \rowcolor[HTML]{EFEFEF} 
    \cellcolor[HTML]{EFEFEF} & screen\_time\_total & 1378 & noti\_removed\_wo\_dups & 220 & screen\_max\_episode & 396 & screen\_time\_total & 225 & app\_not-found & 84 
    \\
    
    \rowcolor[HTML]{EFEFEF} 
    \multirow{-3}{*}{\cellcolor[HTML]{EFEFEF}\textbf{Studying}} & day\_period & 1146 & app\_photography & 178 & screen\_time\_per\_episode & 247 & weekend & 154 & day\_period & 43 \\
    
     & day\_period & 271 & day\_period & 518 & day\_period & 61 & proximity\_std & 38 & app\_not-found & 37 
     \\
     
     & app\_tools & 98 & app\_not-found & 180 & app\_not-found & 26 & proximity\_max & 29 & wifi\_mean-rssi & 23 
     \\
     
    \multirow{-3}{*}{\textbf{Eating}} & app\_not-found & 61 & activity\_still & 72 & app\_video players \& editors & 23 & app\_communication & 18 & wifi\_max-rssi & 21 
    \\
    
    \rowcolor[HTML]{EFEFEF} 
    \cellcolor[HTML]{EFEFEF} & app\_entertainment & 715 & day\_period & 326 & app\_video players \& editors & 397 & app\_entertainment & 151 & wifi\_mean-rssi & 51 
    \\
    
    \rowcolor[HTML]{EFEFEF} 
    \cellcolor[HTML]{EFEFEF} & app\_not-found & 426 & app\_not-found & 325 & wifi\_std\_rssi & 85 & app\_not-found & 59 & app\_lifestyle & 38 
    \\
    
    \rowcolor[HTML]{EFEFEF} 
    \multirow{-3}{*}{\cellcolor[HTML]{EFEFEF}\textbf{\makecell[c]{Watching\\something}}} & weekend & 334 & wifi\_num\_of\_devices & 217 & app\_entertainment & 66 & notifications\_posted & 58 & weekend & 29 \\
    
     & app\_social & 1381 & touch\_events & 503 & wifi\_num\_of\_devices & 112 & app\_tools & 64 & app\_tools & 95 
     \\
     
     & screen\_time\_total & 565 & screen\_time\_total & 355 & wifi\_connected & 93 & app\_causal & 58 & proximity\_max & 58 
     \\
     
    \multirow{-3}{*}{\textbf{\makecell[c]{Online comm./\\Social media}}} & screen\_max\_episode & 473 & app\_not-found & 354 & screen\_time\_total & 92 & screen\_time\_total & 42 & proximity\_mean & 48 \\
    
    \rowcolor[HTML]{EFEFEF} 
    \cellcolor[HTML]{EFEFEF} & weekend & 3167 & day\_period & 455 & weekend & 357 & app\_not-found & 119 & notifications\_posted\_wo\_dups & 148 
    \\
    
    \rowcolor[HTML]{EFEFEF} 
    \cellcolor[HTML]{EFEFEF} & screen\_num\_of\_episodes & 745 & weekend & 289 & day\_period & 260 & notifications\_posted & 104 & weekend & 112 
    \\
    
    \rowcolor[HTML]{EFEFEF} 
    \multirow{-3}{*}{\cellcolor[HTML]{EFEFEF}\textbf{Attending class}} & app\_tools & 476 & app\_not-found & 251 & screen\_max\_episode & 70 & screen\_max\_episode & 37 & screen\_time\_total & 87 
    \\
    
     & steps\_detected & 271 & wifi\_mean\_rssi & 1049 & screen\_time\_per\_episode & 143 & proximity\_mean & 305 & activity\_invehicle & 441 
     \\
     
     & screen\_time\_per\_episode & 210 & wifi\_max\_rssi & 848 & proximity\_mean & 129 & proximity\_max & 304 & wifi\_num\_of\_devices & 226 
     \\
     
    \multirow{-3}{*}{\textbf{Working}} & screen\_num\_of\_episodes & 206 & wifi\_min\_rssi & 633 & screen\_max\_episode & 124 & proximity\_std & 292 & activity\_walking & 163 
    \\
    
    \rowcolor[HTML]{EFEFEF} 
    \cellcolor[HTML]{EFEFEF} & day\_period & 337 & day\_period & 191 & app\_medical & 374 & notifications\_posted & 22 & app\_photography & 145 
    \\
    
    \rowcolor[HTML]{EFEFEF} 
    \cellcolor[HTML]{EFEFEF} & app\_tools & 117 & screen\_time\_total & 89 & app\_arcade & 72 & app\_not-found & 16 & app\_trivia & 64 
    \\
    
    \rowcolor[HTML]{EFEFEF} 
    \multirow{-3}{*}{\cellcolor[HTML]{EFEFEF}\textbf{Resting}} & app\_educational & 66 & screen\_max\_episode & 75 & day\_period & 55 & touch\_events & 14 & app\_maps \& navigation & 23 
    \\
    
     & app\_books \& reference & 955 & app\_not-found & 167 & app\_not-found & 215 & cellular\_lte\_min & 252 & app\_adventure & 21 
     \\
     
     & app\_comics & 93 & touch\_events & 122 & wifi\_std\_rssi & 109 & app\_tools & 83 & app\_comics & 16 
     \\
     
    \multirow{-3}{*}{\textbf{Reading}} & app\_news \& magazines & 93 & day\_period & 121 & wifi\_max\_rssi & 77 & location\_altitude & 76 & location\_altitude & 6 
    \\
    
    \rowcolor[HTML]{EFEFEF} 
    \cellcolor[HTML]{EFEFEF} & activity\_onfoot & 3518 & activity\_onfoot & 1582 & steps\_detected & 376 & steps\_detected & 285 & activity\_walking & 25
    \\
    
    \rowcolor[HTML]{EFEFEF} 
    \cellcolor[HTML]{EFEFEF} & activity\_walking & 3497 & activity\_walking & 1579 & steps\_counter & 314 & activity\_walking & 101 & activity\_onfoot & 25 
    \\
    
    \rowcolor[HTML]{EFEFEF} 
    \multirow{-3}{*}{\cellcolor[HTML]{EFEFEF}\textbf{Walking}} & steps\_detected & 3374 & steps\_detected & 1009 & activity\_walking & 232 & activity\_onfoot & 101 & location\_radius\_of\_gyration & 23 \\
    
     & app\_health \& fitness & 502 & day\_period & 33 & app\_health \& fitness & 931 & app\_health \& fitness & 1248 & wifi\_max\_rssi & 50 
     \\
     
     & day\_period & 233 & wifi\_num\_of\_devices & 32 & proximity\_min & 52 & noti\_removed & 72 & proximity\_std & 41 
     \\
     
    \multirow{-3}{*}{\textbf{Sport}} & notifications\_posted & 132 & wifi\_min\_rssi & 23 & day\_period & 40 & day\_period & 34 & wifi\_mean\_rssi & 41 
    \\
    
    \rowcolor[HTML]{EFEFEF} 
    \cellcolor[HTML]{EFEFEF} & steps\_detected & 283 & activity\_onfoot & 1270 & day\_period & 74 & activity\_walking & 132 & app\_weather & 86 
    \\
    
    \rowcolor[HTML]{EFEFEF} 
    \cellcolor[HTML]{EFEFEF} & activity\_onfoot & 267 & activity\_walking & 1269 & user\_presence\_time & 41 & activity\_onfoot & 131 & app\_auto \& vehicles & 84 
    \\
    
    \rowcolor[HTML]{EFEFEF} 
    \multirow{-3}{*}{\cellcolor[HTML]{EFEFEF}\textbf{Shopping}} & activity\_walking & 265 & steps\_detected & 504 & screen\_num\_of\_episodes & 38 & steps\_detected & 55 & activity\_walking & 79 
    \\
    
    \hline
    \end{tabular}%
    }
\end{table*}

The resulting features across countries for the same activity are different in most cases, highlighting the dataset's diversity and each country's cultural differences or habits. For example, when studying, features regarding screen episodes dominate in the UK, Italy, and Denmark, while the day period appears in Italy, Mongolia, and Paraguay. This could mean that European students tend to use their phones when studying more (or less) than students from Paraguay or Mongolia. This divide is also visible when ``watching something'', which is influenced by the use of entertainment applications in Europe, but not in Paraguay or Mongolia. This effect could be due to the unpopularity of streaming services classified as entertainment applications in the latter two countries, where participants might rely on alternatives. In fact, differences in using streaming services across countries have been studied in prior work, highlighting differences in usage percentages \cite{lotz2021between} and the relations to income level \cite{nhan2020comparison}. On the other hand, it could also be that students watch something on a medium that is not their smartphone. In fact, research shows that young adults aged 18-29 use more online media streaming services as compared to television in the USA \cite{PewResearch2017About}. However, whether similar percentages hold across different countries with contrasting cultures, income levels, and internet quality remains a question. While not conclusive, these could be the reason for entertainment apps not being indicative of ``watching something'' in Mongolia and Paraguay, which are the non-European countries in this study.

For some activities, the top three features are inherent to the nature of the activity. For example, ``reading'' in Italy has features corresponding to reading applications such as books, comics, newspapers, and magazines. Other countries do not show this. The same observation can be made for the ``sports'' activity: health and fitness apps are one of the determining features in European countries. This effect could correspond to participants tracking their workouts using a smartphone app.

The ``walking'' activity has almost the same features in all five countries: steps detected and an on-foot or walking activity detected by the Google Activity Recognition API. This homogeneity is due to the nature of the activity---walking is considered a simple activity. This is also why shopping has some of the same features as walking since participants also walk when they shop. To summarize, in most cases, each country has different defining features when looking at the same activity. For some activities, the features found are inherent to the activity and are usually app categories. Finally, it is worth mentioning that the period of the day is an important feature, which matches what has been observed in Figure \ref{fig:activities-per-hour} --- all activities do not occur at the same frequency throughout the day. 

Finally, it is worth noting that we could expect some of the highly informative features to change over time, with changes to technology use and habits of people, in different countries \cite{xu2022globem, adler2022machine}. For example, a reason for the lack of use of streaming services in certain countries is the lack of laws surrounding the usage of illegally downloaded content (e.g., Germany has strict laws about not using illegal downloads \cite{rump2011kind}). Changes in the laws of countries could change the behavior of young adults. Further, internet prices could also affect the use of streaming services. While bandwidth-based and cheap internet is common in developed countries, it is not the same in developing nations in Asia, Africa, and South America, where internet usage is expensive, hence demotivating streaming. In addition, income levels too could influence captured features a lot. For example, with increasing income levels (usually happens when a country's GDP changes), young adults may use more wearables for fitness tracking, leading to the usage of health and fitness apps on mobile phones. Another aspect that could affect the captured behaviors is the weather condition. All five countries mentioned in this study go through different seasons, as all are somewhat far from the equator. Hence, we could expect changes in features in different seasons. More about this is discussed in the limitations section.
 
\section{Machine Learning-based Inference: Experimental Setup, Models, and Performance Measures}\label{sec:inference}

This study aims to perform a multi-class inference of smartphone sensing data to predict what participants do at a particular time. The input space consists of the features in the tabular dataset previously mentioned. We study the three approaches to the problem as summarized in Figure \ref{fig:overview}, going from country-specific to multi-country.

\begin{figure}
    \includegraphics[width=0.5\textwidth]{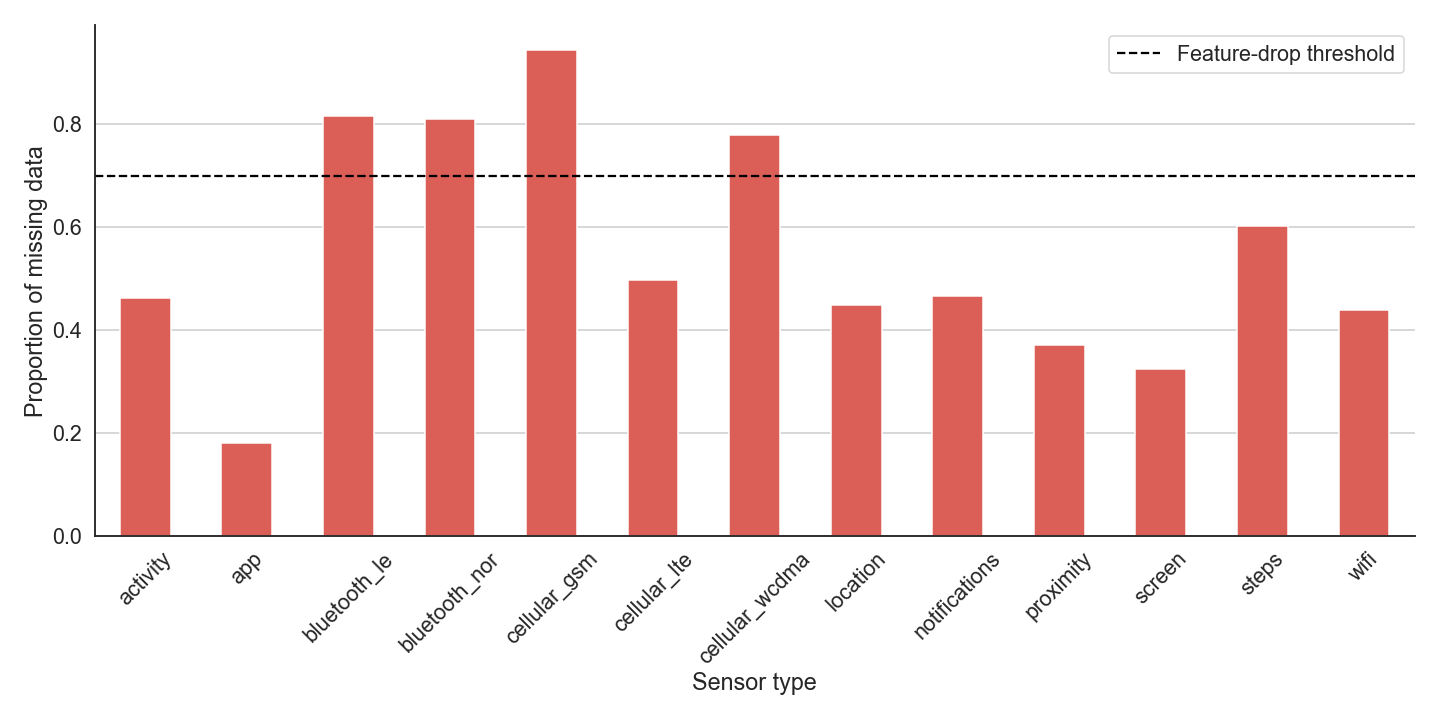}
    \caption{Proportion of missing data per sensor type.}
    \Description{Proportion of missing data per sensor type is shown as a bar chart. More than 90\% of GSM cellular sensor values were unavailable, possibly due to devices being put in airplane mode, sensor failure, or the phone mostly operating with LTE signals. For other modalities, data were available more than 70\% of the time.}
    \label{fig:missing-data}
\end{figure}

\subsection{Data Imputation} The first step in preparing the dataset for inference was data imputation. Missing data in the context of smartphone sensing can occur for multiple reasons \cite{servia-rodriguez_mobile_2017, bae_detecting_2017, meegahapola2022sensing}: the device being on low-consumption mode, the failure of a sensor, or insufficient permissions from the participants. In the dataset we used, we noticed that most sensors have some missing values (see Figure \ref{fig:missing-data}). For example, more than 90\% of GSM cellular sensor values were unavailable, possibly due to devices being put in airplane mode, sensor failure, or the phone mostly operating with LTE signals. To deal with missing values, we decided to drop features from sensors that were missing more than 70\% of their data (refer to the dotted line on Figure \ref{fig:missing-data}) similar to prior work \cite{santani_drinksense_2018}. For the remaining features, and each country individually, we used k-Nearest Neighbour (kNN) imputation \cite{zhang2012nearest} to infer missing information from neighboring samples \footnote{We also tried mean imputation, user-based mean imputation, most frequent value imputation, last observation carried forward (LOCF) imputation, in addition to kNN. However, we obtained the best results for inferences with kNN. In addition, using kNN is common in studies that used passive sensing \cite{zhang2022predicting, rashid2020predicting, zhou2018missing, xu2021understanding}. Hence, we only reported results obtained with kNN.}.

\subsection{Models and Performance Measures}  To conduct all experiments, we used the scikit-learn \cite{scikit-learn} and Keras \cite{chollet2015keras} frameworks, with Python. We first trained the following two baseline models: one that always predicts the most frequent label and another that randomly predicts targets by considering the class distribution. This will allow us to understand if the trained models perform better than a randomized guess. The experiments were carried out with the following model types: Random Forest Classifier \cite{breiman_random_2001} (RF), AdaBoost with Decision Tree Classifier \cite{hastie_multi-class_2009}, and Multi-Layer Perceptron neural networks (MLP) \cite{wang2003artificial} \footnote{We initially tried out other model types such as Gradient Boosting and XGBoost in addition to the reported models. Results for these models were not reported considering their performance and page limits. All these model types are commonly used in small mobile sensing datasets that are in tabular format \cite{merrill2022self, biel_bitesnbits_2018, meegahapola_one_2021}}. The first two inherently leverage class imbalance, and RFs also facilitate the interpretability of results. Each experiment was carried out ten times to account for the effect of randomness. For each experimental setup, we reported the mean and standard deviation across the ten runs for the following metrics: F1 score \cite{F1score2022}, and the area under the Receiver Operating Characteristic curve (AUROC) \cite{AUROC2022}. Even though we calculated the accuracies of models, and while the accuracy is easy to interpret, it might not present a realistic picture in an imbalanced data setting. Hence, we did not include it in the results. The weighted macro F1 score computes metrics for each class and averages them following their support, resulting in a metric that considers label imbalance. Moreover, it takes a significant hit if one of the classes has a lot of false positives. A low F1 score could imply that the classifier has difficulty with rare target classes. The AUROC score measures how well the model can distinguish each activity. It can be understood as an average of F1 scores at different thresholds. We also used a weighted macro version to account for label imbalance.  

Next, we examine results for country-specific, country-agnostic, and multi-country approaches \cite{khwaja_modeling_2019}. Finally, for all three approaches, we examine population-level, and hybrid models that correspond to no and partial personalization, respectively, similar to \cite{meegahapola2023generalization, meegahapola2022sensing, likamwa_moodscope_2013} (training and testing splits were always done with 70:30 ratio):

\begin{figure}
    \includegraphics[width=0.5\textwidth]{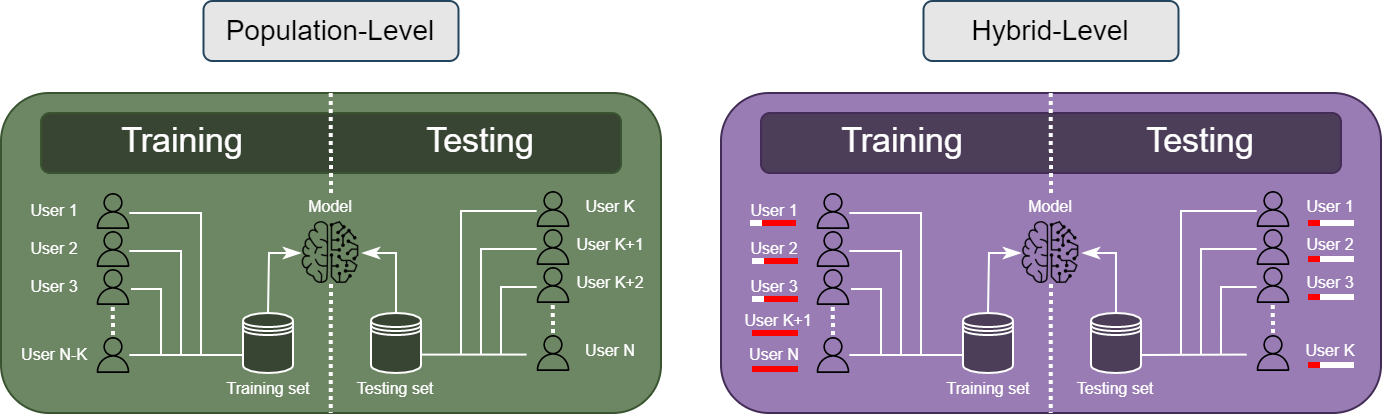}
    \caption{Personalization levels used in the country-specific, country-agnostic, and multi-country approaches. Population-level corresponds to models with no personalization. Hybrid corresponds to models with partial personalization.}
    \label{fig:population-hybrid}
    \Description{Personalization levels used in the country-specific, country-agnostic, and multi-country approaches. Population-level corresponds to models with no personalization. Hybrid corresponds to models with partial personalization.}
\end{figure}

\begin{itemize}[wide, labelwidth=!, labelindent=0pt]

    \item \textbf{Population-Level} model, also known as leave-k-participants-out in country-specific and multi-country approaches, and leave-k-countries-out in country-agnostic approach: the set of participants present in the training set ($\approx$70\%) and the testing set ($\approx$30\%) are disjoint. The splitting was done in a stratified manner, meaning each split was made by preserving the percentage of samples for each class. This represents the case where the model was trained on a subset of the population, and a new set of participants joined a system that runs the model and started using it. 
    \begin{itemize}
        \item In the country-specific approach, this means that data from disjoint participants are in training and testing splits, and everyone is from the same country. E.g., trained with a set of participants in Italy and tested with another set of participants in Italy who were not in the training set. 
        \item In the country-agnostic approach, this means the training set is from one (Phase I) or four (Phase II) countries, and the testing set is from a country not seen in training. E.g., For Phase I --- trained with a set of participants in Italy and tested with a set of participants in Mongolia; Phase II --- trained with a set of participants in Italy, Denmark, UK, and Mongolia, and tested with a set of participants in Paraguay. 
        \item In the multi-country approach, this means a disjoint set of participants in training and testing without considering country information. This is the typical way of training models even when data are collected from multiple countries \cite{servia-rodriguez_mobile_2017}. E.g., trained with a set of participants from all five countries and tested with a set of participants in all five countries who were not in the training set. 
    \end{itemize}
    
    \item \textbf{Hybrid} model, also known as the leave-k-samples-out: the sets of participants in the training and testing splits are not disjoint. Part of the data of some participants present in the testing set ($\approx$70\%) was used in training the models. Testing is done with the rest of the data from the participants ($\approx$30\%). This represents the case where the model was trained on the population, and the same participants whose data were used in training continue to use the model. Hence, models are partially personalized. 
    \begin{itemize}
        \item In the country-specific setting, this means that some data from participants within a country in the testing set can also be in the training set. This represents a scenario where personalization is examined within the country. E.g., trained with a set of participants in Italy and tested with another set of participants in Italy, whose data (70\%) were also used in the training set. The rest of the data (30\%) were used in the testing set.  
        \item In the country-agnostic setting, this means the training set is from one/more countries, and the testing set is from another country, where a percentage of their past data (70\%) was also included in the training. This represents a scenario where personalization is examined when deployed to a new country. E.g., Phase I --- trained with a set of participants in Italy and tested with a set of participants in Mongolia, whose data (70\%) were also used in the training set. Rest of the data (30\%) were used in the testing set; Phase II --- trained with a set of participants in Italy, Denmark, UK, Mongolia, and tested with a set of participants in Paraguay, whose data (70\%) were also used in the training set. The rest of the data (30\%) were used in the testing set. 
        \item In the multi-country setting, this means that training and testing participants are not disjoint, and country information is not considered. This is the typical way of partially personalizing models even when data are collected from multiple countries. E.g., trained with a set of participants from all five countries and tested with a set of participants in all five countries, whose data (70\%) were also used in the training set. The rest of the data (30\%) were used in the testing set. 
    \end{itemize}
    
\end{itemize}

\section{Inference Results}\label{sec:results}

In this section, we present the results of the experiments. First, we discuss results from the country-specific and multi-country approaches, shedding light on \textbf{RQ2}. Then, the country-agnostic approach is discussed by providing answers to \textbf{RQ3} on model generalization.

\subsection{Country-Specific and Multi-Country Approaches (RQ2)}

\paragraph{Country-Specific Approach}

\begin{table*}
    \caption{Mean ($\bar{S}$) and Standard Deviation ($S_\sigma$) of inference
    F1-scores, and AUROC scores computed from ten iterations using
    three different models (and two baselines) for each country separately.
    Results are presented as $\bar{S} (S_\sigma)$, where $S$ is any of the two
    metrics.}
    \label{tab:specific-results}
    \resizebox{\textwidth}{!}{%
    \begin{tabular}{
    >{\columncolor[HTML]{FFFFFF}}l ll
    >{\columncolor[HTML]{FFFFFF}}l 
    >{\columncolor[HTML]{FFFFFF}}l 
    >{\columncolor[HTML]{FFFFFF}}l 
    >{\columncolor[HTML]{FFFFFF}}l 
    >{\columncolor[HTML]{FFFFFF}}l 
    >{\columncolor[HTML]{FFFFFF}}l rr}
     
     &
     \multicolumn{2}{c}{\cellcolor[HTML]{EDEDED}\textbf{Baseline I}} & \multicolumn{2}{c}{\cellcolor[HTML]{EDEDED}\textbf{Baseline II}} & \multicolumn{2}{c}{\cellcolor[HTML]{EDEDED}\textbf{Random Forest}} & 
     \multicolumn{2}{c}{\cellcolor[HTML]{EDEDED}\textbf{AdaBoost}} & \multicolumn{2}{c}{\cellcolor[HTML]{EDEDED}\textbf{MLP}} 
     \\
     
    \multicolumn{11}{c}{\cellcolor[HTML]{FFFFFF}\textbf{Population-Level}} \\
    
    \arrayrulecolor{Gray}
    \hline
    
    & \textit{F1} & \textit{AUROC} & \textit{F1} & \textit{AUROC} & \textit{F1} & \textit{AUROC} & \textit{F1} & \textit{AUROC} & \multicolumn{1}{l}{\textit{F1}} & \multicolumn{1}{l}{\textit{AUROC}} 
    \\
    
    \arrayrulecolor{Gray2}
    \hline
    
    Italy &
    0.17 (0.000) & 0.50 (0.000) &
    0.19 (0.001) & 0.50 (0.001) &
    0.41 (0.001) & 0.71 (0.001) &
    0.39 (0.000) & 0.71 (0.000) & 
    0.38 (0.002) & 0.68 (0.002)
    \\
    
    Mongolia &
    0.26 (0.000) & 0.50 (0.000) &
    0.23 (0.001) & 0.50 (0.001) &
    0.33 (0.002) & 0.62 (0.001) &
    0.33 (0.000) & 0.63 (0.000) &
    0.34 (0.003) & 0.61 (0.004) 
    \\
    
    UK &
    0.17 (0.000) & 0.50 (0.000) &
    0.18 (0.002) & 0.50 (0.001) &
    0.32 (0.004) & 0.63 (0.003) &
    0.31 (0.000) & 0.59 (0.000) &
    0.22 (0.006) & 0.56 (0.003)
    \\
    
    Denmark &
    0.25 (0.000) & 0.50 (0.000) &
    0.24 (0.006) & 0.49 (0.003) &
    0.32 (0.008) & 0.61 (0.006) &
    0.34 (0.000) & 0.57 (0.000) &
    0.25 (0.008) & 0.57 (0.006)
    \\
    
    Paraguay &
    0.19 (0.000) & 0.50 (0.000) &
    0.19 (0.006) & 0.49 (0.002) &
    0.30 (0.004) & 0.59 (0.003) &
    0.28 (0.000) & 0.56 (0.000) &
    0.31 (0.009) & 0.58 (0.004) 
    \\
    
    \arrayrulecolor{Gray}
    \hline
    \multicolumn{11}{c}{\cellcolor[HTML]{FFFFFF}\textbf{Hybrid}} \\
    \arrayrulecolor{Gray}
    \hline
    
    & \textit{F1} & \textit{AUROC} & \textit{F1} & \textit{AUROC} & \textit{F1} & \textit{AUROC} & \textit{F1} & \textit{AUROC} & \multicolumn{1}{l}{\textit{F1}} & \multicolumn{1}{l}{\textit{AUROC}} \\
    
    \arrayrulecolor{Gray2}
    \hline
    
    Italy &
    \multicolumn{1}{r}{0.17 (0.000)} &
    \multicolumn{1}{r}{0.50 (0.000)} & 
    
    \multicolumn{1}{r}{\cellcolor[HTML]{FFFFFF}0.19 (0.001)} & \multicolumn{1}{r}{\cellcolor[HTML]{FFFFFF}0.50 (0.001)} &
    
    \multicolumn{1}{r}{0.63 (0.001)} &
    \multicolumn{1}{r}{0.87 (0.001)} & 
    
    \multicolumn{1}{r}{\cellcolor[HTML]{FFFFFF}0.40 (0.000)} & \multicolumn{1}{r}{\cellcolor[HTML]{FFFFFF}0.73 (0.000)} &
    
    0.51 (0.002) & 0.81 (0.000) 
    
    \\
    Mongolia &
    
    \multicolumn{1}{r}{0.26 (0.000)} & 
    \multicolumn{1}{r}{0.50 (0.000)} &
    
    \multicolumn{1}{r}{\cellcolor[HTML]{FFFFFF}0.23 (0.002)} & \multicolumn{1}{r}{\cellcolor[HTML]{FFFFFF}0.50 (0.001)} &
    
    \multicolumn{1}{r}{0.51 (0.001)} &
    \multicolumn{1}{r}{0.79 (0.001)} &
    
    \multicolumn{1}{r}{\cellcolor[HTML]{FFFFFF}0.34 (0.000)} & \multicolumn{1}{r}{\cellcolor[HTML]{FFFFFF}0.66 (0.000)} &
    
    0.45 (0.002) &
    0.75 (0.002) 
    \\
    
    UK &
    \multicolumn{1}{r}{0.17 (0.000)} &
    \multicolumn{1}{r}{0.50 (0.000)} &
    
    \multicolumn{1}{r}{\cellcolor[HTML]{FFFFFF}0.19 (0.003)} & \multicolumn{1}{r}{\cellcolor[HTML]{FFFFFF}0.50 (0.001)} &
    
    \multicolumn{1}{r}{0.66 (0.001)} &
    \multicolumn{1}{r}{0.88 (0.006)} &
    
    \multicolumn{1}{r}{\cellcolor[HTML]{FFFFFF}0.34 (0.000)} & \multicolumn{1}{r}{\cellcolor[HTML]{FFFFFF}0.68 (0.000)} &
    
    0.58 (0.003) &
    0.83 (0.002) 
    \\
    
    Denmark &
    \multicolumn{1}{r}{0.25 (0.000)} &
    \multicolumn{1}{r}{0.50 (0.000)} &
    
    \multicolumn{1}{r}{\cellcolor[HTML]{FFFFFF}0.24 (0.003)} & \multicolumn{1}{r}{\cellcolor[HTML]{FFFFFF}0.50 (0.002)} &
    
    \multicolumn{1}{r}{0.69 (0.002)} &
    \multicolumn{1}{r}{0.89 (0.001)} &
    
    \multicolumn{1}{r}{\cellcolor[HTML]{FFFFFF}0.41 (0.000)} & \multicolumn{1}{r}{\cellcolor[HTML]{FFFFFF}0.66 (0.000)} &
    
    0.67 (0.002) &
    0.87 (0.002) 
    \\
    
    Paraguay &
    \multicolumn{1}{r}{0.18 (0.000)} &
    \multicolumn{1}{r}{0.50 (0.000)} &

    \multicolumn{1}{r}{\cellcolor[HTML]{FFFFFF}0.19 (0.002)} &
    \multicolumn{1}{r}{\cellcolor[HTML]{FFFFFF}0.49 (0.003)} &
    
    \multicolumn{1}{r}{0.61 (0.003)} &
    \multicolumn{1}{r}{0.84 (0.001)} &

    \multicolumn{1}{r}{\cellcolor[HTML]{FFFFFF}0.30 (0.000)} &
    \multicolumn{1}{r}{\cellcolor[HTML]{FFFFFF}0.61 (0.000)} & 
    
    0.58 (0.002) &
    0.79 (0.001)
    
    \\
    \hline
    \end{tabular}%
    }
\end{table*}

We consider this approach to be the base setting that does leverage country-level diversity in building separate models---each country has its own model independently from others. Table \ref{tab:specific-results} summarizes the results of experiments following the country-specific approach. In the population-level setting, the three models perform more or less similarly, but the RFs are generally better based on F1 and AUROC scores. In the case of the hybrid models, RFs performed the best across the five countries, with AUROC scores in the range of 0.79-0.89, where the lowest was for Mongolia, and the highest was for Denmark. Compared to population-level models, we can notice a substantial bump in performance in the hybrid models, showing the effect of personalization within countries. These results suggest that random forest models applied to a partially personalized setting can recognize complex daily activities from passive sensing data with a good performance. Given this conclusion, even though we got results for all model types for subsequent sections, we will present results only using random forest models. 

\paragraph{Multi-Country Approach}

\begin{figure*}
    \includegraphics[width=\textwidth]{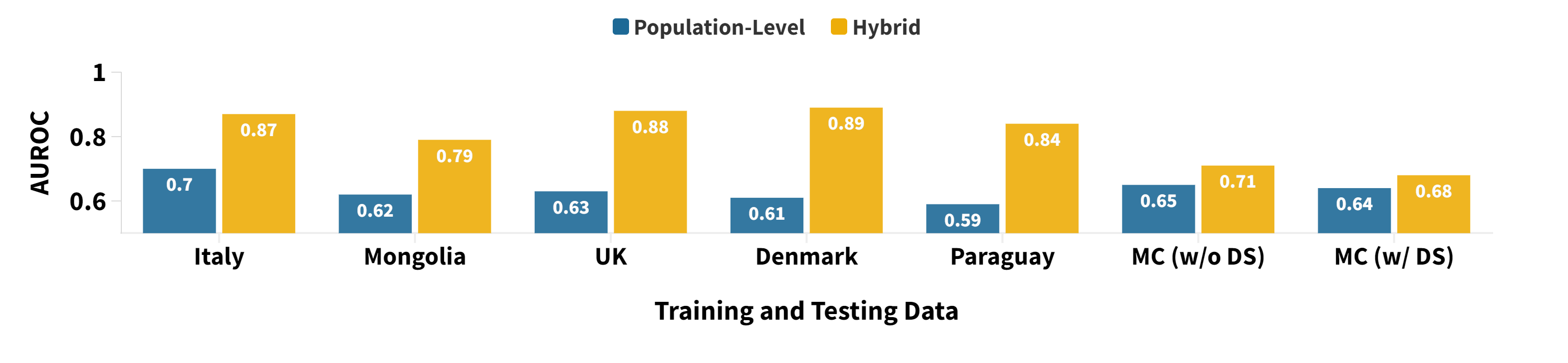}
    \caption{Mean AUROC score comparison for country-specific and multi-country approaches with population-level and hybrid models. MC: Multi-Country; w/o DS: without downsampling; w/ DS: with downsampling.}
    \Description{Diagram shows mean AUROC score comparison for country-specific and multi-country approaches with population-level and hybrid models. MC: Multi-Country; w/o DS: without downsampling; w/ DS: with downsampling. Results show that the country-specific approach provides better results than the multi-country approach, by large margins across both population-level and hybrid models.}
    \label{fig:multi_country_comparison}
\end{figure*}

This approach aims at building a generic multi-country or one-size-fits-all model with the expectation that it would capture the diversity of all countries. All five countries are present in both the training and the testing set. We, therefore, consider all participants of the dataset, regardless of their country, similar to an experiment where country-level diversity is ignored. Hence, we can examine population-level and hybrid models for a multi-country approach in this context. Further, models were evaluated with a dataset with an imbalanced representation from five countries (multi-country w/o downsampling --- MC w/o DS) and a balanced representation from five countries by randomly downsampling from countries with more data to make it equal to the country with the least number of self-reports (i.e., Paraguay) (multi-country w/ downsampling  --- MC w/ DS). The results are shown in Figure~\ref{fig:multi_country_comparison} in comparison to country-specific results. MC w/o DS had an AUROC of 0.71 while MC w/ DS had an AUROC of 0.68, indicating that training on the original data distribution performed better. The reason could in fact be that, more data led to better performance. The expectation of training with downsampled data was to give equal emphasis to each country, expecting that the model would perform well to all countries. However, the result indicates that it is not the case.

These results shed light on our \textbf{RQ2}: learning a multi-country model for complex activity recognition solely using passive smartphone sensing data is difficult (AUROC: 0.709 with hybrid models). It does not yield better performance than the country-specific approach (AUROCs of the range 0.791-0.894). This may stem from the data's imbalance between countries and classes or the context in which the dataset was collected. Another primary reason for this could be behavioral differences in data highlighted in Table~\ref{tab:statistical-analysis}, making it difficult for a model to learn the representation when the diversity of data is unknown. Distributional shifts \footnote{\url{https://huyenchip.com/2022/02/07/data-distribution-shifts-and-monitoring.html\#data-shifts}} across datasets from different countries could be the reason for this. When sensor feature and ground truth distributions (we discussed ground truth distributions in Section~\ref{sec:descriptive}) are different across countries, it could lead to an averaging effect, which would lead to worse-performing models than models for each country. Moreover, it is worth noting that there are not a lot of studies that trained country-specific and multi-country models for performance comparison \cite{phan_mobile_2022}. In one of the only other studies that we found \cite{khwaja_modeling_2019}, personality trait inference performance using smartphone sensor data was better when using country-specific models, similar to what we found for complex daily activity inference. Finally, from a human-centered perspective, recruiting participants to collect smartphone sensing data to build machine learning models means that---rather than targeting large samples from a single country, recruiting a reasonable number of participants from diverse countries could help deploy better-performing models to multiple countries. 

\subsection{Generalization Issues with Country-Agnostic Approach (RQ3)}

We examined this research question with two phases as detailed in Table~\ref{tab:keywords_explaination}. During the first phase, to evaluate the extent to which country-specific models generalize to new countries, we tested models trained with a single country's data in the other four countries separately. In the second phase, to evaluate the extent to which a model trained with four countries generalized to the remaining country, we trained with different combinations of countries and tested on the remaining country.

\begin{figure*}
    \includegraphics[width=\textwidth]{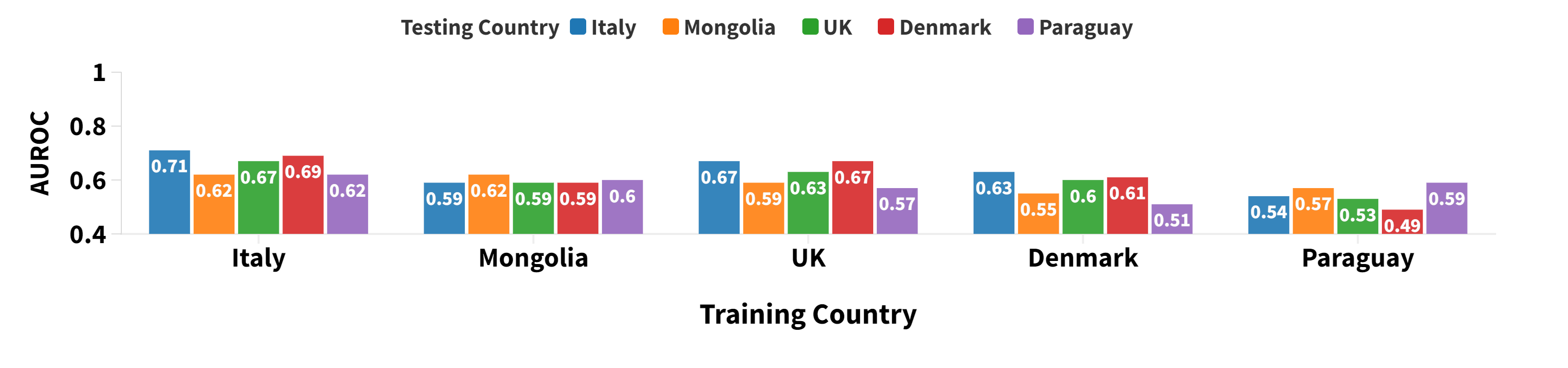}
    \caption{Mean AUROC scores obtained in the country-agnostic approach with population-level models.}
    \label{fig:agnostic_plm}
    \Description{Mean AUROC scores obtained in the country-agnostic approach with population-level models.}
\end{figure*}

\begin{figure*}
    \includegraphics[width=\textwidth]{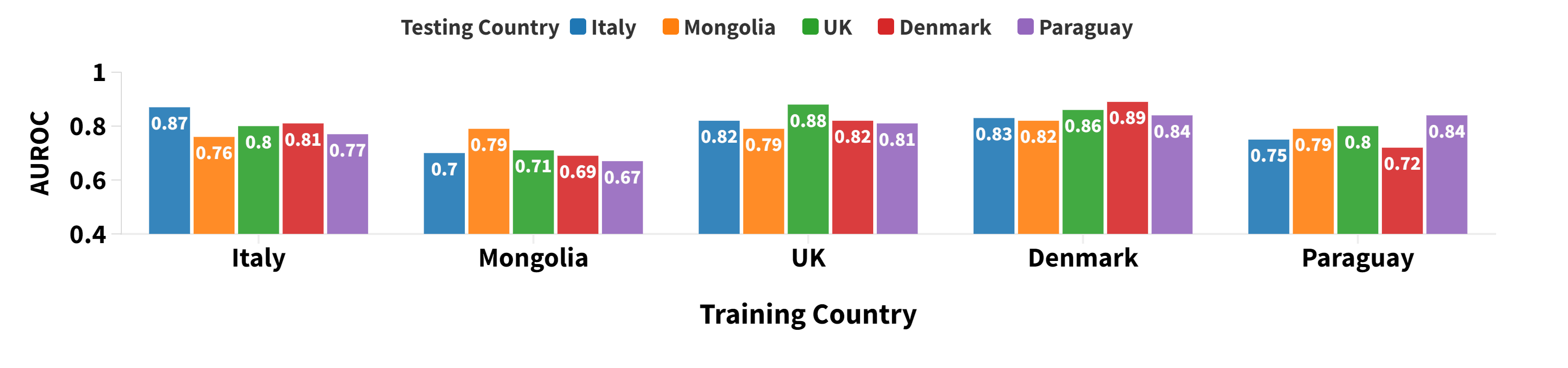}
    \caption{Mean AUROC scores obtained in the country-agnostic approach with hybrid models.}
    \label{fig:agnostic_hlm}
    \Description{Mean AUROC scores obtained in the country-agnostic approach with hybrid models.}
\end{figure*}

\begin{itemize}[wide, labelwidth=!, labelindent=0pt]

    \item \textbf{Phase I:} Figure~\ref{fig:agnostic_plm} summarizes results for population-level models and Figure~\ref{fig:agnostic_hlm} summarizes results for hybrid models. To allow easy comparison, in both figures, the result mentioned as the performance of a country, when tested on the same country is the result from Table~\ref{tab:specific-results}. For instance, at the population-level, Italy had an AUROC of 0.71 according to Table~\ref{tab:specific-results}, and this is marked in Figure~\ref{fig:agnostic_plm} where both Training and Testing country is Italy. Population-level results suggest that the country-agnostic approach tends to perform better in countries geographically close to the country where the model was originally trained. For example, the Italy model had an AUROC of 0.71 for the Italian populations in a population-level setting and performed better in Denmark (AUROC: 0.69) and the UK (AUROC: 0.67) than it did in Mongolia (AUROC: 0.62) or Paraguay (AUROC: 0.62). Similar results can also be observed for hybrid models, where the Italian model performed better in Denmark and UK. This observation suggests that college students from countries within the same geographic region (Europe) could have behaviors that translate to similar smartphone usage and contexts during periods of doing similar activities. This is consistent with the observations made in the descriptive analysis above, where the countries that deviate from the general {trends} are usually those outside Europe. In summary, even after using the same experimental protocol when collecting mobile sensing data, we could still observe a distribution shift of data by the performance of models across geographically distant countries. 
    
    \item \textbf{Phase II: } The second phase looked into extending the work done in phase I. Instead of testing a country-specific model in a new country, we were interested in testing a model already exposed to diverse data (e.g., from four countries) in a new country. We present results for random forest models (because they performed the best across experiments) where the training set consisted of data from four countries, and the testing set had data from the fifth. As suggested in prior studies \cite{khwaja_modeling_2019}, each country contributed equally to the training set in terms of data volume, which means we had to downsample the data from each country to a common count (which was equal to the minimum number of data points available from one country). Table \ref{tab:country-agnostic} presents the results for experiments of the second phase. Similar to previous cases, we observed an increase in performance from population-level to hybrid models. More generally, and by looking at the F1 and AUROC scores, the performance of the hybrid models in the country-agnostic approach is lower than that of the same model in the country-specific approach. This is somewhat expected since including data from other distributions (i.e., other countries) in the training set increases the data's variance and makes it more difficult to represent all distributions accurately. This drop in performance could also be due to the downsampling. For instance, in a model where we train with four countries, including Italy and Paraguay, Italy represents the largest portion of the dataset compared to Paraguay, which is the smallest. When reducing the number of samples in each country to that of Paraguay, a lot of information is lost in the other countries: the larger the original dataset is, the larger the loss gets. This could explain the low performance of country-agnostic models in Italy and Mongolia, especially in the hybrid setting.
    
\end{itemize}

In addition, when comparing different modeling approaches, the results with Multi-Country w/o Downsampling are similar to those found in Phase II (hybrid) of the country-agnostic approach, which was expected since the training sets are similar. However, the bump in performance when going from population-level to hybrid is less noticeable here compared to previous cases. Furthermore, MC w/ DS performs worse than the previous approach, with an AUROC of 0.68 compared to 0.71. This could be because we lose much data from many countries due to downsampling, reducing models' representational ability. To summarize, a hybrid model in a country-agnostic approach can not predict complex activities better than its country-specific counterpart. Furthermore, while more data often means better performances, this does not apply when the data follow different distributions, one per country in this case. This suggests that each country has specific characteristics that make learning one representation difficult.

These results shed light on our \textbf{RQ3}: complex activity recognition models trained in specific countries often generalize reasonably to other countries (especially with hybrid models). However, the performance is not comparable to the country-specific approach, suggesting that there is still a distributional shift between countries. In fact, in Section~\ref{sec:descriptive}, we discussed how the labels used in the inference (i.e., shown in Figure~\ref{fig:activities-per-hour}--complex daily activities such as resting, studying, reading, etc.) had different distributions across the five countries. Further, the extent of the generalization often depended on whether countries are geographically closer (i.e., within Europe) or not. This result is in line with findings from previous studies \cite{phan_mobile_2022, khwaja_modeling_2019} that highlighted the effect of geographic dimensions (i.e., country of data collection) on mobile sensing model performance. For example, \cite{khwaja_modeling_2019} found that country-specific models that used mobile sensing data as input, could perform well for the inference of three personality traits--Extraversion, Agreeableness, and Conscientiousness. Furthermore, we would also like to highlight that the issue regarding distributional shifts and generalization is an open problem in multimodal mobile sensing, as highlighted by two recent studies that examined similar datasets collected from the same country in different time periods~\cite{xu2022globem, adler2022machine}. This is possibly due to behavioral changes over time leading to different distributions in sensor data and ground truth. Our results go beyond this and show that even if data is collected within the same time period and with the same protocol, distributional shifts could still occur due to country differences.

\begin{table}
    \caption{Mean ($\bar{S}$) and Standard Deviation ($S_\sigma$) of 
    F1-scores and AUROC scores obtained by testing each Country-Agnostic model (trained in four countries) on data from a new country. Results are presented as $\bar{S} (S_\sigma)$,
    where $S$ is any of the two metrics.}    
    \label{tab:country-agnostic}
    \resizebox{0.5\textwidth}{!}{%
    \begin{tabular}{
    >{\columncolor[HTML]{FFFFFF}}l l
    >{\columncolor[HTML]{FFFFFF}}l 
    >{\columncolor[HTML]{FFFFFF}}l 
    >{\columncolor[HTML]{FFFFFF}}l }
    
     & \multicolumn{2}{c}{\cellcolor[HTML]{EDEDED}\textbf{Population-Level}} & \multicolumn{2}{c}{\cellcolor[HTML]{EDEDED}\textbf{Hybrid}} 
    \\

    \textit{Test Country} &
    \textit{F1} &
    \textit{AUROC} &
    \textit{F1} &
    \textit{AUROC} 
    \\
    
    \arrayrulecolor{Gray2}
    \hline
    
    Italy &
    0.33 (0.005) & 0.65 (0.006) &
    0.37 (0.004) & 0.71 (0.002)
    \\
    
    Mongolia & 
    0.30 (0.011) & 0.60 (0.004) &
    0.37 (0.006) & 0.67 (0.003) 
    \\
    
    UK &
    0.29 (0.004) & 0.63 (0.005) &
    0.47 (0.004) & 0.78 (0.002)
    \\
    
    Denmark &
    0.38 (0.006) & 0.65 (0.006) &
    0.63 (0.008) & 0.86 (0.004) 
    \\
    
    Paraguay &
    0.28 (0.005) & 0.59 (0.006) &
    0.55 (0.006) & 0.80 (0.008) 
    \\ 
    
    \arrayrulecolor{Gray2}
    \hline
    \end{tabular}}
\end{table}

\begin{figure*}
    \includegraphics[width=0.8\textwidth]{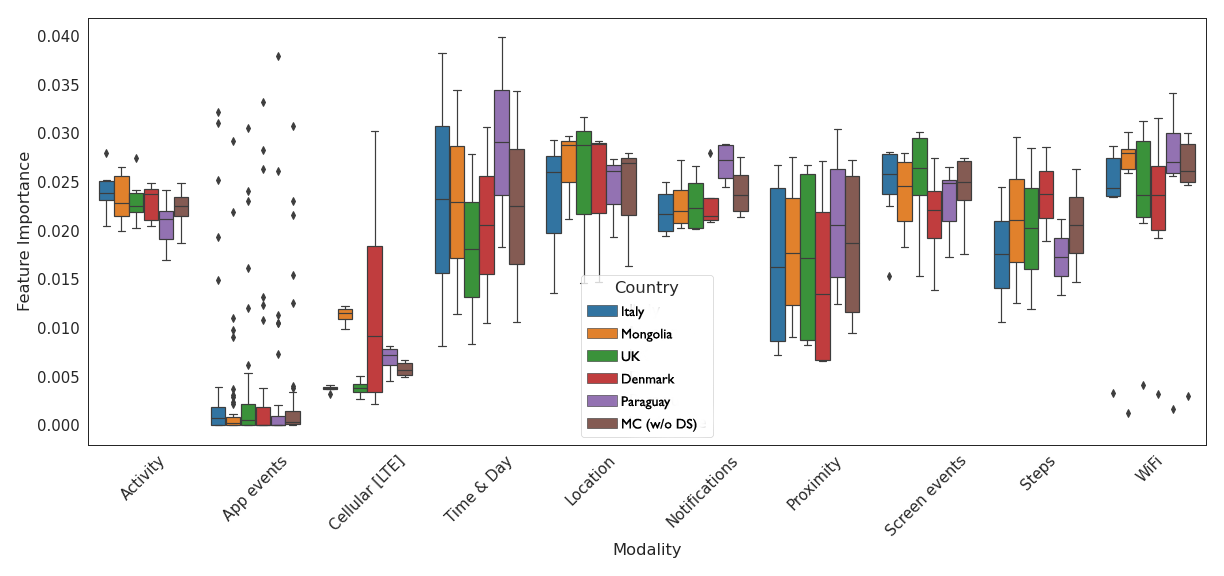}
    \caption{Feature importance of each feature category for hybrid country-specific and multi-country models.}
    \Description{Feature importance of each feature category for hybrid country-specific and multi-country models. These values were obtained using random forest classifiers.}
    \label{fig:feat-importances}
\end{figure*}

\subsection{Feature Importance for Complex Daily Activity Recognition}\label{sec:features}

The random forest models trained in our experiments inherently provide the Gini importance of the features seen during training \cite{breiman_classification_1998}. In Figure \ref{fig:feat-importances}, each set of box plots represents the distribution of feature importances for a given modality (as defined in Table \ref{tab:agg-features}) for hybrid models under the country-specific approach and multi-country approach (MC w/o DS). A first look shows that the multi-country distribution deviates from other countries for all sensing modalities. For example, one cellular feature in the model from Denmark is more important than the other models' cellular features. The temporal, WiFi and notification features are more important in Paraguay than in other countries. App events are mostly unimportant, except for a few outliers across all countries. This is reasonable given that out of the long list of app types used for the analysis, participants frequently used only a few types (e.g., entertainment, social, educational, health and fitness, etc.). Our analysis showed that the outliers here are, in fact, the apps used by participants the most. By looking at the top whiskers of each set of box plots, the most predictive features overall are part of the following modalities: time \& day, wifi, app usage, simple activity type, and location. 
\
\section{Discussion}\label{sec:discussion}

\subsection{Summary of Results}
We examined a multi-country smartphone sensing dataset to develop inference models of complex daily activities. Our primary goal was to seek whether reasonably performing complex daily activity recognition models could be trained using multimodal sensor data from smartphones. Then, our goal was to identify differences among countries visible through smartphone usage and to leverage these differences to decide whether it makes sense to build country-specific or generic multi-country models, and whether models generalize well. We believe these findings are important when designing and deploying sensing and ML-based apps and systems in geographically diverse settings. The main findings for the three research questions can be summarized as follows:
\begin{itemize}[wide, labelwidth=!, labelindent=0pt]
    \item \textbf{RQ1}: Different features in each country can characterize an activity. Their distributions throughout the day also vary between countries and seem to be affected. This finding points towards biases that could get propagated if proper care is not taken during the design and data collection phase of studies involving people and smartphones. In Section \ref{subsubsec:biases}, we discuss this in more detail under a set of biases: construct bias \cite{he2012bias}, sample bias \cite{meegahapola2020smartphone}, device-type bias \cite{blunck2013heterogeneity}, and bias from user practices \cite{van2004bias}.
    
    \item \textbf{RQ2}: It is feasible to train models with the country-specific approach to infer 12 complex activities from smartphone data. Furthermore, personalization within countries increases performance (AUROCs of the range 0.79-0.89). Hence, the country-specific approach outperforms the multi-country approach, which only yields an AUROC of 0.71 with hybrid models. However, building multi-country models solely from sensing features is a non-trivial task that might require more effort with regard to data balance and feature selection. Our results also show that the sedentary lifestyles of the pandemic world can be captured with country-specific partially personalized machine learning models. In addition, we also show that multimodal smartphone sensors could be used to recognize complex daily activities that go beyond binary inferences to 12-class inferences. In Section \ref{subsubsec:inthewild}, we discuss why real-life studies are important to capture complex emerging lifestyles; in Section \ref{subsubsec:contextaware}, we also discuss how complex daily activities could be useful to design novel context-aware applications.
    
    \item \textbf{RQ3}: Under the country-agnostic approach, we found that models generalize reasonably to new countries. However, unsurprisingly, the performance is not as high as when the model was tested in the same country where it was trained. Interestingly, models trained in European countries performed better in other European countries than in Paraguay or Mongolia. This issue broadly falls under the topic of domain shifts, which remains under-explored in mobile sensing literature. We elaborate more on this in Section~\ref{subsubsec:domainshift}.
    
\end{itemize}

\subsection{Implications}\label{subsec:implications}

Our work has implications aligned to both theoretical and practical aspects.

\subsubsection{Accounting for Country Biases in Study Design (RQ1).}\label{subsubsec:biases} Studies using sensing data drawn from geographically diverse samples (i.e., different countries) should account for and understand the \emph{sources of biases} that can occur at different stages of the study. Our study, and also previous studies on human behavior, sociology, and psychology, allow an understanding of these aspects in detail. For example, the following taxonomy can be used to characterize such biases \cite{phan_mobile_2022}. \textit{(i) Construct bias} occurs when the target is expressed differently across countries, depending on countries’ norms or environmental factors \cite{he2012bias}. For example, the ``walking'' activity in one country where physical exercise is not widespread could be labeled as ``walking'', whereas in a country where it is an activity done for fitness by many people, so it could be labeled as a ``sport'' as well. Hence, some behaviors can be specific to a particular environment or group of people. \textit{(ii) Sample bias} concerns the comparability of diverse samples that can be impacted by the recruitment process in each country \cite{meegahapola2020smartphone}. For example, if the samples in each country differ in age or gender, sensing data would likely not have similar distributions across countries. \textit{(iii) Device-type bias} is due to the differences in the devices used by participants across countries and in environmental factors affecting sensor measurements \cite{blunck2013heterogeneity}. Devices worldwide are not equipped with the same software and hardware, and similar sensors can differ in accuracy and precision (e.g., Apple devices are more prominent in developed countries, whereas Android phones are common in others). Finally, the \textit{(iv) bias from user practices} arises when participants from different countries use their mobile phones differently \cite{van2004bias}. Examples abound: how a phone is physically carried could distort measurements; how sensors are disabled to save battery or mobile data (especially in countries where unlimited mobile data plans are not standard) also changes what is measured; and different motivations to use certain apps in different countries also changes the resulting logs \cite{lim2014investigating}. Phan et al. \cite{phan_mobile_2022} have proposed a set of mitigation strategies that aim to reduce biases and foster fair approaches to diversity-aware research. To achieve these objectives, the authors recommend taking several steps during both the planning and implementation phases of the study. During the planning phase, researchers are advised to acquire knowledge about potential cross-country differences and relevant environmental factors, with the assistance of local informants. Furthermore, researchers should ensure that their study targets are comparable across countries and that they exist in each country being studied. In the implementation phase, the authors suggest inclusive recruitment strategies that aim to make each sampled country representative of a given target. These recommended strategies are important in promoting diversity-aware research and mitigating the potential for biases that can skew results.

\subsubsection{Activities Captured in Real-Life Studies (RQ2).}\label{subsubsec:inthewild} In terms of theoretical implications, it is worth highlighting that the set of activities that we considered are complex behaviors that can not be typically captured during in-lab studies. Fine-grained sensing-based activity recognition studies help increase performance on simple activities (e.g., walking, running, sitting, climbing stairs, etc. --- that have a repetitive nature in sensor data) that can be captured in in-lab settings. In contrast, building sensing-based ML models to capture complex daily behaviors requires conducting real-life studies. Activities like studying, attending classes, or shopping is hard to replicate in lab settings. Further, while simple activities might not have led to differences in model performances across countries, complex daily activities tightly bound with cultural, country-level, or geographic norms lead to differences in behaviors, leading to differences in the sensed data. In this context, prior work in the domain has not focused on this aspect enough, in our view. Even more so, we believe that studies must capture data from diverse communities to build models that work for all intended users. While this is a challenging task, it is much needed for the field of research to mature for more real-life use cases. Our study is one of the first studies in this direction.

\subsubsection{Novel Applications of Context-Aware Mobile Apps (RQ2).}\label{subsubsec:contextaware} In terms of practical implications, our findings point towards adding context awareness to mobile applications. Current mobile applications provide context-aware services, interventions, and notifications based on location and simple activity sensing \cite{mehrotra2017intelligent, meegahapola2020smartphone}. However, a range of potential applications that go beyond the current offering could become feasible with complex daily activity recognition. For example, previously, a smartphone would only know that a user was sitting in a particular place. With complex activity recognition, it would know that a user is studying, attending a lecture, reading, or eating, which all entail sitting. For example, if the student is reading, studying, or attending a lecture, automatically putting the phone in silent or do-not-disturb mode might make sense, even though, in many cases, people forget to do so. In summary, complex daily activity recognition could offer diverse use cases to build mobile applications around in the future.

\subsubsection{Domain Adaptation for Multimodal Mobile Sensing (RQ3).}\label{subsubsec:domainshift} Another theoretical implication can be described in a machine learning sense. We discussed the challenges of generalization and domain shifts in our smartphone sensor dataset. We described how this shift affects model performance, specifically for complex daily activity recognition with multimodal sensors. Although biases, distributional shifts, and model generalization have been widely studied in other domains such as natural language processing \cite{elsahar2019annotate}, speech \cite{sun2017unsupervised}, and computer vision \cite{luo2019taking}, smartphone sensing studies have yet to receive sufficient attention \cite{gong2021dapper}. We demonstrated that model personalization (hybrid setting) could reduce distributional shifts to a certain extent. In a way, according to transfer learning-related terms, this approach is similar to fine-tuning an already trained model for a specific user to achieve model personalization \cite{chen2020fedhealth}. Such strategies for personalization have been used in prior work \cite{meegahapola_one_2021}. However, recent research in domain adaptation has shown limitations in mobile sensing, particularly with regard to time series data \cite{wilson2022domain}. The diversity of wearable device positioning poses a persistent issue in human activity recognition, which affects the performance of recognition models \cite{chang2020systematic, mathur2019unsupervised}. Wilson et al. \cite{wilson2022domain} conducted a study of domain adaptation in datasets captured from individuals of different age groups, yet the findings are limited to simpler time series accelerometer data. Other works admit that the current lack of solutions for domain adaptation and generalization from smartphone and wearable data presents an opportunity for future exploration \cite{adler2022machine, xu2022globem}. We have added to the literature by confirming that domain adaptation techniques are necessary for multi-country, multimodal smartphone sensor data. In addition, even on a fundamental level, approaches that allow quantifying cross-dataset distributional differences for multimodal sensing features and target labels (e.g., activity, mood, social context, etc.) separately, are lacking in the domain. Research on such aspects could allow us to better understand distributional shifts in sensor data, to better counter it with domain adaptation techniques in multimodal settings.

\subsection{Limitations}

While the dataset covers five different countries from three
continents, students’ behavior in other countries and
continents could differ from what we have already
encountered. In addition, even though we found geographically closer countries performing well in Europe, such findings need to be confirmed for other regions where geographically closer countries could have contrasting behaviors and norms (e.g., India and China). Furthermore, the weather conditions in different countries during the time period of data collection could be slightly different. All five countries mentioned in this study go through different seasons, as all are somewhat far from the equator. Hence, we could expect changes in features in different seasons. However, in practical terms, collecting data in similar weather conditions is not feasible.

When aggregating sensor data around self-reports, the data corresponding to the moment the participant was filling out the self-report is considered a part of the activity he/she was doing at the time. This noise could alter the recognition task if the window's size is small enough. However, even though this could affect results if we intended to increase model performance in a fine-grained sensing task, we do not believe this noise affects the results significantly regarding our findings on diversity awareness. In addition, it is worth noting that the way we model our approach with a tabular dataset is similar to prior ubicomp/mobile sensing studies done in real life \cite{meegahapola2020smartphone} because we do not have continuous ground truth labels. Hence, it restrains us from modeling the task as a time-series problem, which is how a majority of activity recognition studies \cite{straczkiewicz2021systematic} with continuous accurate ground truth measurements follow. So, the results should be interpreted with the study's exploratory nature in mind.

Further, it is worth noting that we could expect some of the highly informative features used in models to change over time, with changes to technology use and habits of people, in different countries \cite{xu2022globem, adler2022machine}. For example, a reason for the lack of use of streaming services in certain countries (discussed in Section \ref{sec:descriptive}) is the lack of laws surrounding the usage of illegally downloaded content (e.g., Germany has strict laws about not using illegal downloads \cite{rump2011kind}). Changes in the laws of countries could change the behavior of young adults. Further, internet prices could also affect the use of streaming services. While bandwidth-based and cheap internet is common in developed countries, it is not the same in developing nations in Asia, Africa, and South America, where internet usage is expensive, hence demotivating streaming. In addition, income levels too could influence captured features a lot. For example, with increasing income levels (usually happens when a country's GDP changes), young adults may use more wearables for fitness tracking, leading to the usage of health and fitness apps on mobile phones.

The amount of data for each country is highly imbalanced. For a fair representation of each country, having the same number of participants and self-reports per country would ensure that a classifier learns to distinguish classes from each country equally. However, Italy and Mongolia are dominant in the current state of the dataset. If not done carefully, down-sampling would result in a loss of expressiveness and variance, making it difficult to discern different classes in a multi-country approach. Another imbalance is found among class labels, where activities such as sleeping or studying are more frequent than others. However, this does make sense since we do not expect all activities to appear at the same frequency in a participant's day or week. Further, we reported F1 and AUROC scores that are preferred in such imbalanced settings. 

Finally, the dataset was collected in November 2020, during the Covid-19 pandemic, when most students stayed home due to work/study-from-home restrictions. This explains why most of the relevant features found in the statistical analysis are screen events and app events. While some relevant features are relative to proximity and WiFi sensors, there are very few regarding activity and location unless the activity corresponds to physical activities. This is probably an effect of a context where movements were highly discouraged. From another perspective, the behavior of college students from all countries during this time period reflects remote work or study arrangements. We could expect these practices to continue for years as more universities and companies adopt remote work/study culture. Hence, while many prior studies in ubicomp used phone usage features and sensing features for activity/behavior/psychological trait inference tasks, our findings indicate that phone usage features could be even more critical in the future with remote study/work settings due to sedentary behavior, that would limit the informativeness of sensors such as location and inertial sensors. 

\subsection{Future Work}

The study's population for the dataset collection consisted of students. Therefore, it might be worth exploring how people from different age groups use their smartphones and how their daily behavior is expressed through that usage. In addition to visible diversity, it is known that deep diversity attributes (innate to humans and not visible) such as personality (captured with Big Five Inventory \cite{donnellan2006mini}), values (captured with basic values survey \cite{gouveia2014functional} and human values survey \cite{schwartz1994there, schwartz2001extending}), and intelligence (captured with multiple intelligence scale \cite{tirri2008identification}) could also affect smartphone sensor data and activities performed by people \cite{schelenz_theory_2021, khwaja_modeling_2019}. Hence, investigating how such diversity attributes could affect smartphone-based inference models on complex activities, and other target variables, is worth investigating. Further, future work could investigate how the classification performance is affected when excluding the sensing data corresponding to the time taken to fill the self-report about activities by participants. Finally, domain adaptation for multi-modal smartphone sensor data across time and countries, remains an important problem worth investigating in future work. 
\section{Conclusion}\label{sec:conclusion}

In this study, we examined the daily behavior of 637 students in Italy, Mongolia, the United Kingdom, Denmark, and Paraguay using over 216K self-reports and passive sensing data collected from their smartphones. The main goal of this study was to, first examine whether multimodal smartphone sensor data could be used to infer complex daily activities, which in turn would be useful for context-aware applications. Then, to examine whether models generalize well to different countries. We have a few primary findings: \textit{(i)} While each country has its day distribution of activities, we can observe similarities between the geographically closer countries in Europe. Moreover, features such as the time of the day or the week, screen events, and app usage events are indicative of most daily activities; \textit{(ii)} 12 complex daily activities can be recognized in a country-specific and personalized setting, using passive sensing features with reasonable performance. However, extending this to a multi-country model does not perform well, compared to the country-specific setting; and \textit{(iii)} Models do not generalize well to other countries (at least compared to within-country performance), and especially to geographically distant ones. More studies are needed along these lines regarding complex daily activity recognition and also other target variables (e.g., mood, stress, fatigue, eating behavior, drinking behavior, social context inference, etc.), to confirm the findings. Hence, we believe research around geographic diversity awareness is fundamental for advancing mobile sensing and human behavior understanding for more real-world utility across diverse countries. From a study design sense, we advocate the idea of collecting data from diverse regions and populations to build better-represented machine learning models. From a machine learning sense, we advocate the idea of developing domain adaptation techniques to better handle multimodal mobile sensing data collected from diverse countries.

\begin{acks}
This work was funded by the European Union’s Horizon 2020 WeNet
project, under grant agreement 823783. We deeply thank all
the volunteers across the world for their participation in the study.
\end{acks}

\bibliographystyle{ACM-Reference-Format}
\bibliography{sections/9_bibliography}

\end{document}